\documentclass[aps,twocolumn,nofootinbib,showpacs,prd,aps,10pt]{revtex4}
\usepackage[dvips]{graphicx}
\usepackage[english]{babel}
\usepackage[T1]{fontenc}
\usepackage{mathrsfs}
\usepackage[tbtags]{amsmath}
\usepackage{amssymb}
\usepackage{amsxtra}
\usepackage{amsopn}
\usepackage{latexsym}
\usepackage[mathcal]{eucal}
\usepackage{mathtools}

\newcommand{\BE}{\begin{equation}}
\newcommand{\EE}{\end{equation}}
\newcommand{\BA}{\begin{align}}
\newcommand{\EA}{\end{align}}
\newcommand{\Tr}{\mathrm Tr}
\newcommand{\nn}{\nonumber}

\newcommand{\kkki}{ \frac{i{\rm d}^4k}{(2\pi)^4}}
\newcommand{\qqqi}{ \frac{i{\rm d}^4q}{(2\pi)^4}}
\newcommand{\kkkE}{ \frac{{\rm d}^4k_E}{(2\pi)^4}}

\newcommand{\kkEd}{ \frac{{\rm d}^dk_E}{(2\pi)^d}}

\newcommand{\fsl}[1]{\ensuremath{\mathrlap{\!\not{\phantom{#1}}}#1}}
\renewcommand{\Re}{\mathop{\rm Re}}
\renewcommand{\Im}{\mathop{\rm Im}}

\begin{document}

\title{Analytic structure of QCD propagators in Minkowski space}

\author{Fabio Siringo}

\affiliation{Dipartimento di Fisica e Astronomia 
dell'Universit\`a di Catania,\\ 
INFN Sezione di Catania,
Via S.Sofia 64, I-95123 Catania, Italy}

\date{\today}

\begin{abstract}
Analytical functions for the propagators of QCD, including a set of chiral quarks, are derived
by a one-loop massive expansion in the Landau gauge, deep in the infrared. By analytic continuation,
the spectral functions are studied in Minkowski space, yielding a direct proof of positivity violation
and confinement from first principles.The dynamical breaking of chiral symmetry is described on the same footing
of gluon mass generation, providing a unified picture. While dealing with the exact Lagrangian, the expansion
is based on {\it massive} free-particle propagators, is safe in the infrared and is equivalent to the 
standard perturbation theory in the UV. By dimensional regularization, all diverging mass terms cancel exactly  
without including mass counterterms that would spoil the gauge and chiral symmetry of the Lagrangian. 
Universal scaling properties are predicted for the inverse dressing functions and shown to be satisfied by
the lattice data. Complex conjugated poles are found for the gluon propagator, in agreement with the 
{\it i-particle} scenario.
\end{abstract}




\maketitle

\maketitle

\section{introduction}

While we still miss a fully consistent analytical method for the study of QCD in the infrared (IR),
important progress has been achieved in recent years and a general consensus has been reached
on a decoupling scenario with a finite gluon propagator and a dynamically generated gluon mass\cite{cornwall} 
in the Landau gauge.

Most of the non-perturbative approaches that have been developed so far rely on numerical calculations in the
Euclidean space, where synergic studies 
by lattice simulations\cite{bogolubsky,twoloop,dudal,binosi12,burgio15,bowman04,bowman05,su2glu,su2gho}, 
Schwinger-Dyson 
equations\cite{aguilar8,aguilar10,aguilar14,aguilar14b,papa15,papa15b,huber14,huber15g,huber15b,fischer2009,fischer2003}
and variational methods\cite{reinhardt04,reinhardt05,reinhardt08,reinhardt14,sigma,
sigma2,gep2,varqed,varqcd,genself,ptqcd0} 
have drawn a clear picture for the propagators of QCD deep in the IR.

However, since physics happens in Minkowski space, much important dynamical information cannot be
extracted by the Euclidean formalism, unless we have an analytic function that can be continued to
the physical space or the whole numerical analysis is carried out in Minkowski space\cite{straussDSE}.
Even the concept of a dynamical mass has no obvious meaning for confined particles
like gluons and quarks. Thus, it is still unknown if the propagators have poles on the real axis, while
some evidence of positivity violation has been demonstrated by indirect arguments.

As a matter of fact, the analytic continuation of a limited set of data points is an ill-defined problem
and any numerical attempt would only give qualitative results at best. Nevertheless, by a linear
regularization strategy, a K\"allen-Lehmann spectral function was reconstructed in Ref.\cite{dudal14}
from the lattice data of the gluon propagator, giving some direct evidence for positivity violation and 
the absence of any discrete mass pole on the physical real axis.

Important insights also come from physically motivated phenomenological models,  providing simplified
analytical descriptions that can be easily continued to 
Minkowski space\cite{shakinPRD,iparticle,tissier10,tissier11,tissier14,sorella15,dudal15,dudal08}.
For instance, the existence of complex poles was predicted by the refined version\cite{dudal08,dudal08b,dudal11}
of the Gribov-Zwanziger model\cite{GZ}.

Quite recently, an analytical approach has been proposed that is based on a different expansion point for
the exact Lagrangian of pure Yang-Mills theory in the Landau gauge\cite{ptqcd,ptqcd2}. 
The new expansion is around a massive free-particle 
propagator, yielding a {\it massive} loop expansion with massive particles in the internal lines of the
Feynman graphs. From first principles, without adding spurious counterterms or phenomenological parameters, 
at one-loop the expansion provides analytical universal functions for the dressing functions, 
predicting some scaling properties that are satisfied by the data of lattice simulations. 
In the Euclidean space and Landau gauge, the massive expansion 
is in impressive agreement with the lattice data\cite{scaling}
and the one-loop propagators are analytic functions that can be easily continued and studied in Minkowski space.

In this paper, the massive expansion is extended to full QCD by the inclusion of a set of
chiral quarks in the Lagrangian. The dynamical breaking of chiral symmetry is described on the same footing
of gluon mass generation, providing a unified picture from first principles.
Analytic functions for the one-loop propagators are derived in the Landau gauge
and continued to Minkowski space where the K\"allen-Lehmann spectral functions are studied in detail.
Thus, the paper has the twofold aim of providing an analytical and consistent framework for the study of QCD 
in the infrared and of disclosing the analytic properties of the propagators in Minkowski space.

By including a set of chiral quarks, the massive expansion provides an analytical approach to QCD from
first principles, without any need for mass counterterms or other phenomenological parameters for
describing the dynamical generation of mass. All mass divergences cancel exactly, yielding a finite
and consistent picture of the dynamical breaking of chiral symmetry together with the dynamical
generation of the gluon mass. However, in its present development, the method does not allow a prediction
of the ratio between the two mass scales and is sensitive to the choice of the subtraction point. These
are the only free parameters in the calculation and the sensitivity to these parameters is
a measure of the accuracy of the one-loop approximation. In fact, having used the exact Lagrangian, any deviation
from the exact result can only arise from the truncation of the expansion. Thus, while based on a
perturbative expansion, the method has a variational nature, with an accuracy that can be increased by
tuning the mass ratio and the subtraction point to optimal values that minimize the effect of higher loops.
In that sense, the method can be seen as a special case of the {\it optimized perturbation theory} that has been
discussed by many authors in the past\cite{stevenson,stevensonRS,stevenson16} and has been recently improved by
RG methods\cite{KN1,KN2,KN3,KN}.

A fully consistent variational estimate of the optimal parameters would require the calculation of an
effective potential\cite{stevenson,var,light,bubble} 
or some real observable quantity, which is out of the aim of the present paper. 
On the other hand, 
by a comparison with the data of lattice simulations, we find that the weight of higher-loop corrections 
can be made very small by a judicious choice of masses and renormalization constants. Thus, with the aim
of describing the analytic structure of the propagators, in the present paper we adopt the strategy of
optimizing the unknown parameters by a direct comparison with the lattice data in the Euclidean space.
Once optimized, the one-loop propagators are continued to Minkowski space, yielding a detailed and
direct description of the analytic properties and spectral functions. Of course, we just assume that, 
being negligible in the Euclidean space, the neglected higher-loop corrections are still small in Minkowski space.

The study of the spectral functions gives direct predictions on the dynamics of quarks and gluons.
For instance, no poles are found for the gluon on the physical positive real axis and the positivity constraints
are strongly violated for quarks and gluons, as expected for confined degrees of freedom.
Moreover, complex conjugated poles are found for the gluon propagator, 
confirming the {\it i-particle} scenario\cite{iparticle}
predicted by the refined version\cite{dudal08,dudal08b,dudal11} of the Gribov-Zwanziger model\cite{GZ}.

The paper is organized as follows:
in Section II the massive expansion of Ref.\cite{ptqcd2} is extended to chiral QCD in the Landau gauge;
in Section III, the case of pure Yang-Mills theory is examined, the propagators are continued to Minkowski space 
and their spectral density is studied in detail; in Section IV, the full chiral QCD is discussed, a dynamical
mass is obtained for the quarks and compared with the lattice data in the Euclidean space,
the spectral functions are derived for ghosts, gluons and quarks; in Section V the main results of the paper
are discussed. The explicit analytical expressions of the propagators are derived in the Appendix.

\section{The Massive Expansion in the Chiral Limit}

In this section we include a set of chiral quarks in the massive expansion that has been recently 
developed for pure Yang Mills theory\cite{ptqcd0,ptqcd,ptqcd2}. While the inclusion of 
bare masses for the quarks would be straightforward, we prefer to deal with the chiral limit in this paper
for several reasons. First of all, we show that the expansion can describe the breaking of chiral
symmetry on the same footing as dynamical mass generation for gluons. Even without bare masses, all 
diverging mass terms are canceled exactly at 
one loop with no need of mass counterterms.
On the other hand, the inclusion of bare masses would give a phenomenological model, with a set of renormalized
masses to be fixed by the phenomenology, without adding too much to the quark sector of other models
with massive gluons that have been already studied\cite{tissier14}. Those models were shown to
be in good agreement with the data of lattice simulations by a proper choice of the parameters.
Moreover, most of the mass of the constituent quarks arises from the interaction, and it seems to
be more challenging to reproduce the data of the simulations in the chiral limit without the aid
of any parameter, from first principles.

The full Lagrangian of QCD, including $N_f$ massless chiral quarks, can be written as
\BE
{\cal L}_{QCD}={\cal L}_{YM}+{\cal L}_{fix}+{\cal L}_{FP}+{\cal L}_{q}
\label{LQCD}
\EE
where ${\cal L}_{YM}$ is the Yang-Mills term
\BE
{\cal L}_{YM}=-\frac{1}{2} \Tr\left(  \hat F_{\mu\nu}\hat F^{\mu\nu}\right)
\EE
${\cal L}_{fix}$ is a gauge fixing term, ${\cal L}_{FP}$ is the ghost Lagrangian
arising from the Faddev-Popov determinant and ${\cal L}_{q}$ is the quark Lagrangian
\BE
{\cal L}_{q}=\sum_{i=1}^{N_f}
\bar\Psi_i\left[i\fsl{\partial}-g{\ensuremath{\mathrlap{\>\not{\phantom{A}}}A_a}}\hat T_a\right]\Psi_i.
\label{Lq}
\EE
In terms of the gauge fields, the tensor operator $\hat F_{\mu\nu}$ is 
\BE
\hat F_{\mu\nu}=\partial_\mu \hat A_\nu-\partial_\nu \hat A_\mu
-i g \left[\hat A_\mu, \hat A_\nu\right]
\EE
where
\BE
\hat A^\mu=\sum_{a} \hat T_a A_a^\mu
\EE
and the generators of $SU(N)$ satisfy the algebra
\BE
\left[ \hat T_a, \hat T_b\right]= i f_{abc} \hat T_c
\EE
with the structure constants normalized according to
\BE
f_{abc} f_{dbc}= N\delta_{ad}.
\label{ff}
\EE
If a generic covariant gauge-fixing term is chosen
\BE
{\cal L}_{fix}=-\frac{1}{\xi} \Tr\left[(\partial_\mu \hat A^\mu)(\partial_\nu \hat A^\nu)\right]
\EE
the total action can be written as ${\cal S}_{tot}={\cal S}_0+{\cal S}_I$ where the free-particle term is
\begin{align}
{\cal S}_0&=\frac{1}{2}\int A_{a\mu}(x)\delta_{ab}\> {\Delta_0^{-1}\>}^{\mu\nu}(x,y)\> A_{b\nu}(y) 
{\rm d}^dx{\rm d}^dy \nn \\
&+\sum_{i=1}^{N_f}\int
\bar\Psi_i(x)\>S_0^{-1} (x,y) \>\Psi_i(y){\rm d}^dx{\rm d}^dy
\nn \\
&+\int \omega^\star_a(x) \delta_{ab}\>{{\cal G}_0^{-1}}(x,y) \>\omega_b (y) {\rm d}^dx{\rm d}^dy
\label{S0}
\end{align}
and the interaction is
\BE
{\cal S}_I=\int{\rm d}^dx \left[ {\cal L}_{qg}+ {\cal L}_{3g} +   {\cal L}_{4g}+{\cal L}_{gh} \right]
\label{SI}
\EE
with the four local interaction terms that read
\begin{align}
{\cal L}_{qg}&=-g\sum_{i=1}^{N_f} \bar\Psi_i \>{\ensuremath{\mathrlap{\>\not{\phantom{A}}}A_a}}\hat T_a\>\Psi_i\nn\\
{\cal L}_{3g}&=-g  f_{abc} (\partial_\mu A_{a\nu}) A_b^\mu A_c^\nu\nn\\
{\cal L}_{4g}&=-\frac{1}{4}g^2 f_{abc} f_{ade} A_{b\mu} A_{c\nu} A_d^\mu A_e^\nu\nn\\
{\cal L}_{gh}&=-g f_{abc} (\partial_\mu \omega^\star_a)\omega_b A_c^\mu.
\label{Lint}
\end{align}
In Eq.(\ref{S0}), $\Delta_0$, $S_0$ and ${\cal G}_0$ are the standard free-particle propagators for
gluons, quarks and ghosts and their Fourier transforms are
\begin{align}
{\Delta_0}^{\mu\nu} (p)&=\Delta_0(p)\left[t^{\mu\nu}(p)  
+\xi \ell^{\mu\nu}(p) \right]\nn\\
\Delta_0(p)&=\frac{1}{-p^2},\qquad  S_0 (p)=\frac{1}{\fsl{p}}, \qquad {{\cal G}_0} (p)=\frac{1}{p^2}.
\label{D0}
\end{align}
Here the transverse and longitudinal projectors are defined as
\BE
t_{\mu\nu} (p)=\eta_{\mu\nu}  - \frac{p_\mu p_\nu}{p^2};\quad
\ell_{\mu\nu} (p)=\frac{p_\mu p_\nu}{p^2}
\label{tl}
\EE
where $\eta_{\mu\nu}$ is the metric tensor. 

As shown in Ref.\cite{ptqcd2} a shift of the pole in the propagators can be introduced by an unconventional
splitting of the total action. We may add and subtract the arbitrary terms $\delta {\cal S}_g$,
$\delta {\cal S}_q$ in the total action
\begin{align}
{\cal S}_0&\rightarrow {\cal S}_0+\delta {\cal S}_q+\delta {\cal S}_g\nn\\
{\cal S}_I&\rightarrow {\cal S}_I-\delta {\cal S}_q-\delta {\cal S}_g
\label{shift}
\end{align}
and take
\begin{align}
\delta {\cal S}_g&= \frac{1}{2}\int A_{a\mu}(x)\>\delta_{ab}\> \delta\Gamma_g^{\mu\nu}(x,y)\>
A_{b\nu}(y) {\rm d}^dx{\rm d}^dy \nn\\
\delta {\cal S}_q&=\sum_{i=1}^{N_f}\int
\bar\Psi_i(x)\>\delta\Gamma_q (x,y)\>\Psi_i(y){\rm d}^dx{\rm d}^dy
\label{dS}
\end{align}
where the vertex functions $\delta\Gamma_g$, $\delta\Gamma_q$  are given by
a shift of the inverse propagators
\begin{align}
\delta \Gamma_g^{\mu\nu}(x,y)&=
\left[{\Delta_m^{-1}}^{\mu\nu}(x,y)- {\Delta_0^{-1}}^{\mu\nu}(x,y)\right]\nn\\
\delta \Gamma_q(x,y)&=\left[{S_M^{-1}} (x,y)- {S_0^{-1}} (x,y)\right]
\label{dG}
\end{align}
and ${\Delta_m}^{\mu\nu}$, $S_M$ are massive free-particle propagators 
\begin{align}
{\Delta_m^{-1}}^{\mu\nu} (p)&={\Delta_m}(p)^{-1} t^{\mu\nu}(p)  
+\frac{-p^2}{\xi}\ell^{\mu\nu}(p)\nn\\
{\Delta_m}(p)^{-1}&=-p^2+m^2,\qquad S_M (p)^{-1}=\fsl{p}-M.
\label{Deltam}
\end{align}
Here the  masses $m$ and $M$ are totally arbitrary. Since $\delta {\cal S}_q$ and $\delta {\cal S}_g$ 
are added and subtracted again, the total action cannot depend on the masses, but any expansion in powers 
of the new shifted interaction ${\cal S}_I \to {\cal S}_I-\delta {\cal S}_q-\delta {\cal S}_g$ is going 
to depend on them at any finite order because of the truncation. 
Thus, while we are not changing the content of the  theory, the emerging perturbative approximation
is going to depend on the masses and can be optimized by a choice of $m$ and $M$ that minimizes the
effects of higher orders, yielding a variational tool disguised to look like a perturbative method\cite{ptqcd2}.
The idea is not new and goes back to the works on the Gaussian effective
potential\cite{stevenson,su2,LR,HT,stancu2,stancu,superc1,superc2,AF}
where an unknown mass parameter was inserted in the zeroth order propagator and subtracted
from the interaction, yielding a pure variational approximation
with the mass that acts as a variational parameter.

The shifts $\delta {\cal S}_q$, $\delta {\cal S}_g$ have two effects on the resulting perturbative expansion: 
the free-particle propagators are replaced by massive propagators and  new two-point vertices 
are added to the interaction, arising from the counterterms that read 
\BE
\delta \Gamma_g ^{\mu\nu}(p)= m^2 t^{\mu \nu} (p), \qquad \delta \Gamma_q (p)= -M.
\EE

The Landau gauge is the optimal choice for the massive expansion
since transverse and longitudinal Lorentz sub-spaces do not mix. Although other covariant
gauges could be explored, their study would require the numerical solution of integral equations\cite{ptqcd2}.
From now on we will take the limit $\xi\to 0$. 
In Eq.(\ref{Deltam}) the gluon propagator becomes transverse and we can simplify the notation 
and drop the projectors $t^{\mu\nu}$ everywhere whenever each term is transverse. 
Moreover we drop all color indices in the diagonal matrices.

We can use the standard formalism of Feynman graphs with massive zeroth
order propagators $\Delta_m$, $S_M$ and the counterterms $\delta \Gamma_g=m^2$, $\delta \Gamma_q=-M$
that must be added to the standard vertices of QCD in Eq.(\ref{Lint}).

Since the resulting total interaction is a mixture of terms that depend on 
the coupling strength $g$ and counterterms that do not vanish in the limit $g\to 0$,
a perturbative expansion in powers of the total interaction would contain at any order different
powers of $g$ but the same number of vertices (including the counterterms among vertices) and 
we may define the order of a term as the number of vertices in the graph.

Assuming that the effective coupling never reaches values that are too large\cite{ptqcd2}, 
we may neglect higher loops and take a double expansion in powers of the total 
interaction and in powers of the coupling, retaining graphs with $n$ vertices
at most and no more than $\ell$ loops. 

\begin{figure}[b] 
\centering
\includegraphics[width=0.22\textwidth,angle=-90]{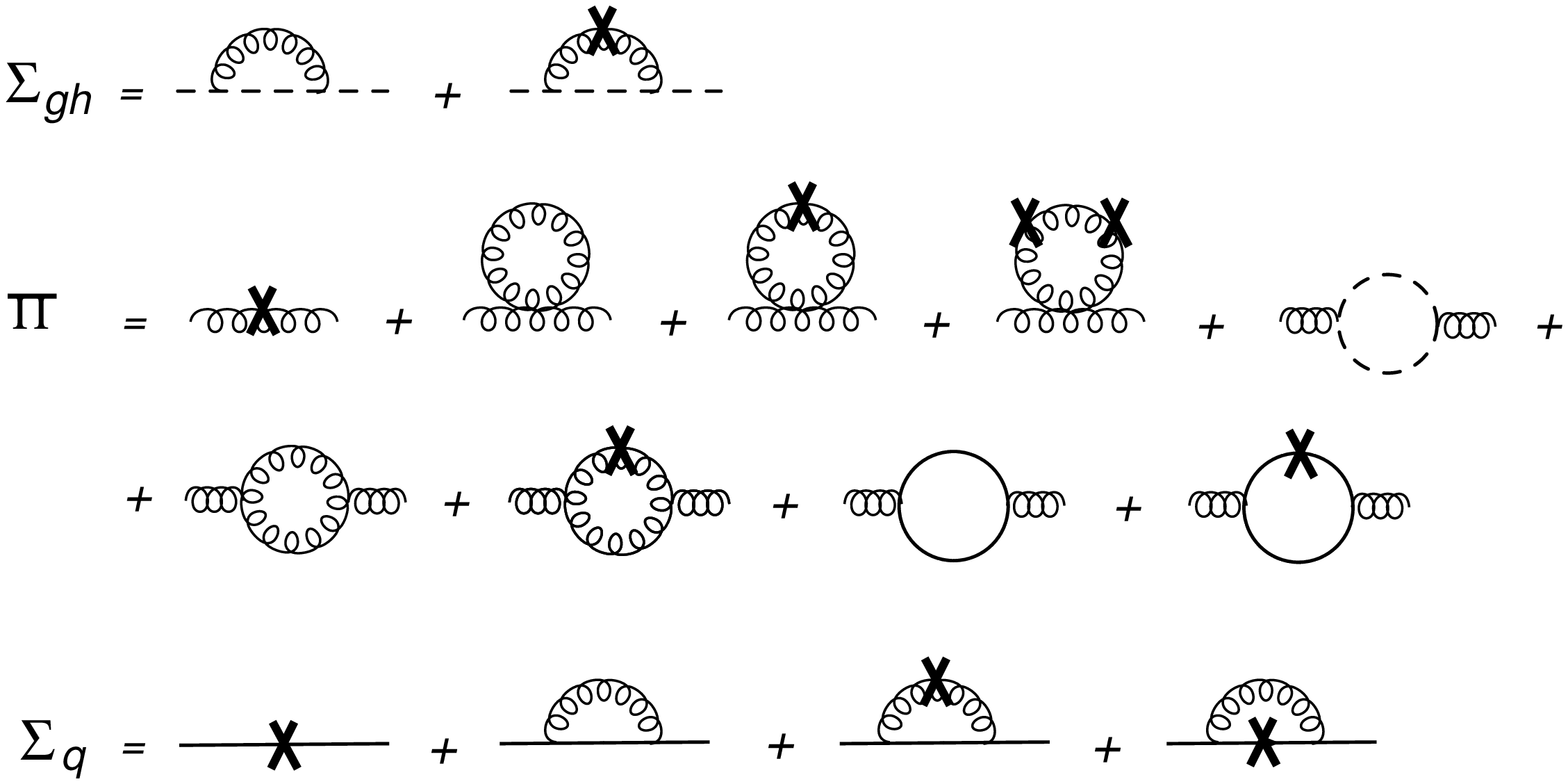}
\caption{Two-point graphs with no more than three vertices and no more than one loop. 
The crosses are the counterterms $\delta \Gamma_g=m^2$, $\delta \Gamma_q=-M$.
In this paper, the quark and ghost self energy and the gluon polarization  
are obtained by the sum of all the graphs in the figure.}
\label{F1}
\end{figure}

The graphs contributing to the quark and ghost self-energy and to the gluon polarization are shown
in Fig.\ref{F1} up to the third order and one-loop. 

We observe that if a counterterm (a cross in Fig.\ref{F1}) is inserted in an $n$-order $\ell$-loop graph,
the number of loops does not change but the order increases by one, giving  an  $(n+1)$-order $\ell$-loop
{\it crossed} graph. At one-loop, the somehow arbitrary truncation of the expansion, at a given order, 
can be shown to have no important effects on the outcome of the calculation. In fact, while higher-order terms
would add very small corrections in the UV because of the factor $(-p^2+m^2)^{-n}$, they might be relevant 
in the limit $p\to 0$ where we find a hierarchy in the significance  of the crossed graphs.
The most important effect arises at tree-level since the tree-graphs in Fig.\ref{F1} cancel
the entire shift of the pole in the propagators. Thus, a finite mass can only arise from loops
and the massive expansion does not predict any mass for the photon in QED. 
At one-loop, a first insertion of the counterterms gives  diverging crossed
graphs that cancel the mass divergences in the loops entirely. Inclusion of those terms is crucial
for the renormalization of the theory. On the other hand, the insertion of $n$ counterterms in a loop,
with $n\ge 2$, gives finite terms that only add some finite corrections to the graph. Moreover
these corrections are absorbed in part by a change of the renormalization constants without affecting
the final result. The convergence of the expansion is very slow, so that no dramatic changes are observed
if higher order graphs are inserted above the third order. 
In that sense, the minimal choice of a third order expansion has nothing special in itself and 
we will limit ourselves to truncating the expansion at that order and at one-loop, 
which is equivalent to the sum of all the graphs in Fig.\ref{F1}.

An alternative way to recover the same expansion is by taking the standard loop-expansion of QCD and
inserting the geometric expansions of the massless propagators in powers of massive propagators according
to
\begin{align}
\Delta_0 (p)&=\Delta_m(p)\>\sum_{n=0}^{\infty}\left[ m^2 \Delta_m(p)\right]^n\nn\\
S_0(p)&=S_M(p)\>\sum_{n=0}^{\infty}\left[-M S_M(p)\right]^n
\label{geometric}
\end{align}
which is just the Dyson sum of all reducible tree-graphs that cancel the pole shift in the propagators.
That shows how truncating the expansion can lead to nontrivial differences between the massive and the
standard expansion since they would differ by an infinite set of graphs that may contain non-perturbative
effects. Moreover, Eq.(\ref{geometric}) makes it obvious that the content of the exact theory has not changed
but the approximation differs because of the different truncation of the total expansion.

Eq.(\ref{geometric}) provides a simple way of calculating the crossed graphs. Using the explicit
form of the propagators in Eq.(\ref{Deltam}) the geometric expansion in Eq.(\ref{geometric}) can
be written as
\begin{align}
\Delta_0 (p)&=\sum_{n=0}^{\infty} \frac{(-m^2)^n}{n!}\frac{\partial^n}{\partial (m^2)^n}  \Delta_m(p)\nn\\
&=\lim_{\bar m\to m}\exp\left( -\bar m^2\frac{\partial}{\partial m^2}\right)\>\Delta_m(p)\nn\\
S_0(p)&=\sum_{n=0}^{\infty}  \frac{(-M)^n}{n!}\frac{\partial^n}{\partial M^n} S_M(p)\nn\\
&=\lim_{\bar M\to M}\exp\left( -\bar M \frac{\partial}{\partial M}\right)\>S_M(p).
\label{geoexp}
\end{align}
Then, exploiting the formal properties of the exponentials, in the standard massless expansion of QCD,
any $\ell$-loop graph $\Sigma^\ell$ for the self-energy or the polarization (containing the product 
of gluon and quark massless propagators) gives rise to the double expansion with massive crossed graphs
\BE
\Sigma^\ell (p)=\sum_{n=0}^{\infty} \frac{1}{n!}
\left[-m^2\frac{\partial}{\partial m^2} -M \frac{\partial}{\partial M}\right]^n
\>\Sigma_{(m,M)}^\ell (p)
\label{crossedsum}
\EE
where $\Sigma_{(m,M)}^\ell (p)$ is the same standard (uncrossed) $\ell$-loop graph 
but with the massless propagators replaced by the massive ones, $\Delta_0\to \Delta_m$ and $S_0\to S_M$.  
The $n$th term of the sum in Eq.(\ref{crossedsum}) gives the sum of all graphs with $n$ counterterm insertions (crossed $n$ times) that arise from the uncrossed $\ell$-loop graph $\Sigma_{(m,M)}^\ell (p)$.
Eq.(\ref{crossedsum}) provides a simple way to derive the massive double expansion from the standard loop
expansion of QCD. Of course, we do not want to sum Eq.(\ref{crossedsum}) exactly, as its truncation gives
a different approximation with respect to the standard perturbative method.

As discussed in Ref.\cite{ptqcd2}, even if the total Lagrangian has not been changed, the truncation 
spoils its exact symmetries. The quadratic part is not BRST and chiral invariant so that the Slavnov-Taylor 
and Ward identities cannot be satisfied exactly at
any finite order.
However, since the total Lagrangian is still symmetric, the deviations can be made very small by inclusion of
higher order terms and would be vanishing if all terms could be summed up.
Thus the issue becomes a question of convergence of the expansion. 
Optimization by variation of the regularization scheme\cite{stevensonRS,stevenson16} has been proven to be a valid tool 
for achieving a quick convergence of the expansion, yielding reliable results even at one-loop.

A remarkable example of the role played by BRST and chiral invariance is provided by 
the regularization of mass divergences.
In the standard perturbative expansion, the invariance properties of the Lagrangian do not allow any diverging mass
term to come out from the loops. When the quadratic part of the Lagrangian is not invariant and contains bare masses,
spurious mass terms arise at any finite order and their divergence requires the insertion of new counterterms.
However, if the total Lagrangian has not been changed and is still invariant, the divergences must be canceled 
by the new crossed graphs that arise from the modified interaction.
Since the insertion of a counterterm in a loop lowers the degree of divergence of the graph, all spurious mass 
divergences can be canceled at a finite order, yielding a regularized expansion in the IR.

\section{Pure Yang-Mills theory}

Without quarks, the dressed propagators of pure SU(N) Yang-Mills theory
can be written as
\begin{align}
\Delta(p)^{-1}&=-p^2+\frac{5}{8}\alpha m^2-\left[ \Pi(p)-\Pi(0)\right]\nn\\
{\cal G} (p)^{-1}&=p^2-\Sigma_{gh} (p),
\label{dressed}
\end{align}
where the ghost self-energy $\Sigma_{gh}$ and the gluon polarization $\Pi$
were evaluated in Ref.\cite{ptqcd2} as a sum of the graphs in Fig.\ref{F1} 
(omitting quark loops) and
the coupling $\alpha$ is defined as
\BE
\alpha=\frac{3N}{4\pi} \alpha_s;\qquad \alpha_s=\frac{g^2}{4\pi}.
\EE 
The one-loop gluon and ghost propagators are made finite by
standard wave function renormalization. In the
$\overline{MS}$ scheme, setting $d=4-\epsilon$, the gluon wave-function renormalization constant $Z_A$
and the ghost wave-function renormalization constant $Z_\omega$ read
\begin{align}
Z_A&=1+\frac{13\alpha}{9\epsilon}=1+\frac{13}{3}\frac{g^2 N}{16\pi^2}\frac{1}{\epsilon}\nn\\
Z_\omega&=1+\frac{\alpha}{2\epsilon}=1+\frac{3}{2}\frac{g^2 N}{16\pi^2}\frac{1}{\epsilon},
\label{Z}
\end{align}
and agree with the standard one-loop approximation in the UV.

It is useful to introduce the adimensional ghost and gluon dressing functions 
\BE
\chi (p)=p^2{\cal G}(p);\quad J(p)=-p^2 \Delta (p).
\label{dress1}
\EE
Once renormalized by the constants $Z_A$, $Z_\omega$, they are finite and read
\begin{align}
\chi (s)^{-1}&=1+\alpha\left[G(s)-\frac{1}{4}\log{\frac{\mu^2}{m^2}}+{\rm const.}\right]\nn\\
J (s)^{-1}&=1+\alpha\left[F(s)-\frac{13}{18}\log{\frac{\mu^2}{m^2}}+{\rm const.}\right]
\label{dress2}
\end{align}
where $s=-p^2/m^2$ is the Euclidean momentum, $\mu$ is the renormalization scale and the
two adimensional functions $F(s)$, $G(s)$ follow from the sum of polarization and self energy graphs
in Fig.\ref{F1}, respectively, and are reported in Appendix A.
Their asymptotic behaviour is just what we need
for canceling the dependence on $m$ in the dressing functions. In fact, in the UV,  Eq.(\ref{dress2})
can be written as
\begin{align}
\chi (p)^{-1}&=\chi (\mu)^{-1}+\frac{\alpha}{4}\log\frac{-p^2}{\mu^2}\nn\\
J (p)^{-1}&=J (\mu)^{-1}+\frac{13\>\alpha}{18}\log\frac{-p^2}{\mu^2}
\label{dressUV}
\end{align}
which is the standard UV behaviour that we expected by inspection of the renormalization constants Eq.(\ref{Z}).

In units of $m$, the arbitrary choice of the mass parameter is reflected by a slight dependence 
on the renormalization scale $\mu/m$ which is the only other energy scale in the theory.
We can put all the constants together and recast the dressing functions in Eq.(\ref{dress2}) as
\begin{align}
\left[\alpha\> \chi (s)\right]^{-1}&=G(s)+G_0\nn\\
\left[\alpha\> J (s)\right]^{-1}&=F(s)+F_0
\label{dress3}
\end{align}
where the arbitrary choice of the ratio $\mu/ m$ is now reflected by a dependence on the additive 
renormalization constants $F_0$ and $G_0$. While the dependence on $N$ and on the coupling $\alpha_s$
is totally absorbed by a multiplicative renormalization of the dressing functions in Eq.(\ref{dress3}),
the dependence on the additive renormalization constants $F_0$, $G_0$ is not totally compensated by
a change of the renormalization. That is a consequence of the one-loop approximation, since if all higher-order
terms were included the result would not depend on $m$. 
Actually, at one-loop, if the dressing functions are multiplied
by an arbitrary factor $Z=1+\alpha \>\delta Z\approx 1$, that factor should be compensated by the subtraction
of $\alpha \>\delta Z$ on the right-hand sides of Eqs.(\ref{dress2}). Thus an only partial degree of 
compensation reflects a sensitivity to the choice of the scale $m$.
On the other hand, the one-loop approximation
can be optimized by taking additive constants that minimize the effects of higher orders.
That would be equivalent to fixing the best mass ratio $m/\mu$, yielding a sort of variational approximation.
Being equivalent to a variation of the subtraction point $\mu$, any change of the additive constant can
be seen as a variation of the renormalization scheme yielding a special case of 
{\it optimized perturbation theory} that has been proven to be very effective for the convergence 
of the expansion\cite{stevensonRS,stevenson16}.

Eq.(\ref{dress3}) has a strong predictive power. 
A very important consequence is that, up to an arbitrary
{\it multiplicative} renormalization constant, the inverse dressing functions are given by
the universal functions $F(s)$ and $G(s)$ up to an {\it additive} renormalization constant. In other
words, the first derivatives of the inverse functions are fixed and do not depend on any parameter but
on the energy units. Actually, that is a useful property for determining the energy units by a comparison with
the lattice data. Such scaling property is satisfied quite
well by the lattice data for SU(2) and SU(3) that collapse on the same universal curves $F(s)$, $G(s)$ in the
infrared, as shown in Refs.\cite{ptqcd,ptqcd2,scaling}. 
That scaling property confirms that higher order terms can
be made negligible by an optimized choice of the constants $F_0$, $G_0$.

While several strategies can be devised for a variational estimate of the best additive constants $F_0$, $G_0$,
in this paper we explore a different approach. By an appropriate choice of the additive constants, the one-loop 
dressing functions in Eq.(\ref{dress3}) are known to give a very accurate description of the data of 
lattice simulations in the Euclidean space\cite{ptqcd,ptqcd2}. Thus, once the constants have been fixed, 
the analytical functions $F(s)$, $G(s)$ can be continued to Minkowski space, 
providing new important information that can be hardly extracted by the data points of the simulations.
Some attempts at reconstructing an analytic function from the lattice data 
have been reported recently\cite{dudal14}, reaching a good qualitative agreement for some sample functions
and showing interesting insights into the spectral functions. Assuming that the weight of
higher-order terms would remain small even for $s<0$, as it is for $s>0$, the present approach would provide a 
simple way to reach quantitative predictions in the physical space.

\begin{figure}[b] 
\centering
\includegraphics[width=0.35\textwidth,angle=-90]{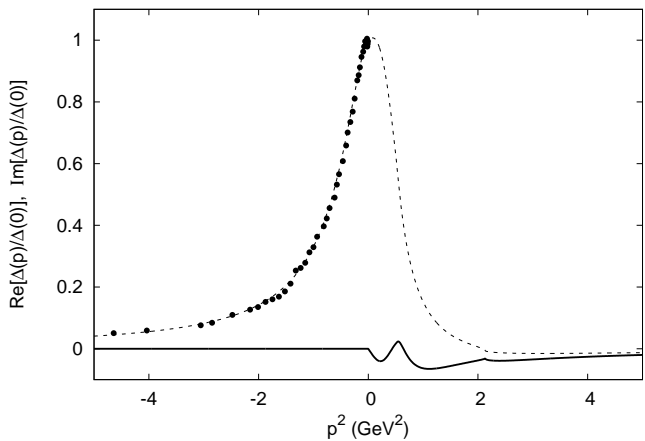}
\caption{The real part (broken line) and the imaginary part (solid line) of the gluon propagator
are displayed together with the lattice data of Ref.\cite{bogolubsky} ($N=3$, $\beta=5.7$, $L=96$). 
The propagator is normalized by its finite value at $p^2=0$ and is evaluated by Eq.(\ref{dress3})
with $F_0=-1.05$ and $m=0.73$ GeV.}
\label{F2}
\end{figure}

\begin{figure}[t] 
\centering
\includegraphics[width=0.35\textwidth,angle=-90]{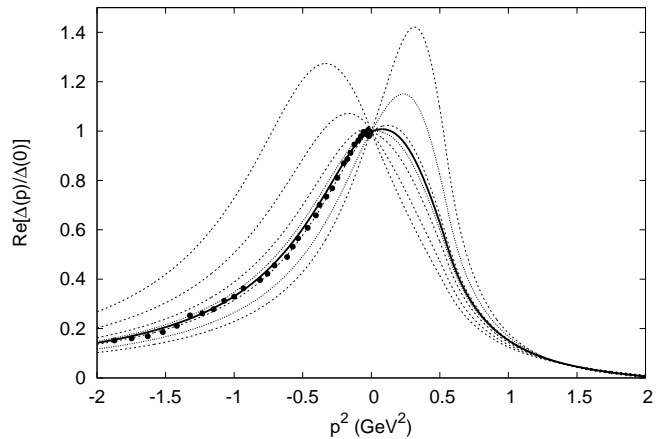}
\caption{Real part of the gluon propagator according to Eq.(\ref{dress3}) for $m=0.73$ GeV and
several values of $F_0$ in the range $-1.6<F_0<-0.6$. The points are the lattice data 
of Ref.\cite{bogolubsky} ($N=3$, $\beta=5.7$, $L=96$). The solid line is obtained for $F_0=-1.05$.}
\label{F3}
\end{figure}

\begin{figure}[t] 
\centering
\includegraphics[width=0.35\textwidth,angle=-90]{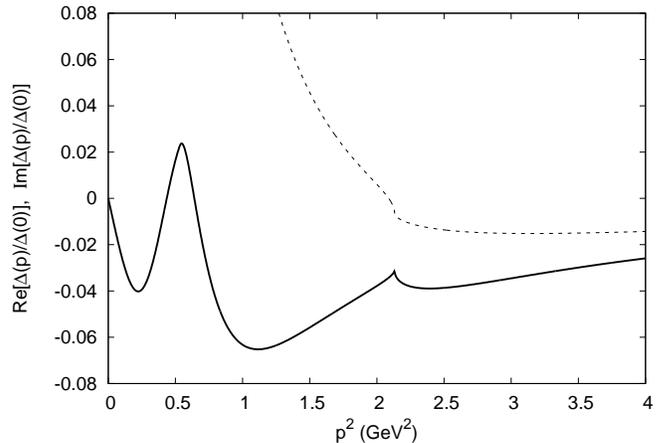}
\caption{The real part (broken line) and the imaginary part (solid line) of the gluon propagator
(enlargement of Fig.2).}
\label{F4}
\end{figure}

For $SU(3)$ and $-p^2<4$ GeV$\>^2$ the lattice data of Ref.\cite{bogolubsky} are very well reproduced by setting
$F_0=-1.05$ and $m=0.73$ GeV in Eq.(\ref{dress3}). 
Some deviation occurs for $-p^2> 4$ GeV$\>^2$ because of the large logs that
require a resummation by RG equations in the UV. 
While that is hardly a problem in the IR, some caution must be taken when attempting to compare with
exact asymptotic results in the UV, like Oehme-Zimmermann superconvergence
relations\cite{OZ}. As shown in Ref.\cite{tissier11}, 
when RG effects are included the massive expansion
is in perfect agreement with the standard perturbative expansion in the UV, since the mass terms become
negligible in the high energy limit. With that limitation,
the gluon propagator can be continued to Minkowski space by setting
$s=-p^2/m^2-i\varepsilon$ and the resulting complex function is shown in Fig.\ref{F2}.

As shown in Fig.\ref{F3}, the propagator is quite sensitive to the choice of $F_0$, allowing a quite
precise estimate of the best value $F_0=-1.05$ that minimizes the higher-order terms. It is remarkable that
the best value occurs very closely to the point where the real part of the propagator seems to be symmetric
in the infrared. Actually, the maximum of the real part is stationary and takes its minimum when the
curve (the solid line in Fig.\ref{F3}) is on the points of the lattice data. That could be relevant for devising
a variational procedure independently from any lattice data. In that respect, exploring the timelike
range $p^2>0$ seems to be important for identifying stationary properties that might be hidden in the
Euclidean space.

Once the constant $F_0$ has been fixed at its best value, the real and imaginary part can be studied in
Minkowski space. 
The imaginary part of the gluon propagator $\Delta$ has a cut for $p^2>0$ and a spectral function 
$\rho(p^2)$ can be defined 
\BE
\rho(p^2)= \frac{1}{\pi} \Im\Delta(p^2+i\varepsilon)
\label{rho}
\EE
which is proportional to the imaginary part above the cut.
According to K\"allen-Lehmann representation, the spectral function must be positive if the particle
is observed among the outgoing states of a scattering process. Thus, confinement requires that
the spectral function $\rho(p^2)$ violate the positivity condition. The imaginary part of $\Delta$
is shown in more detail in Fig.\ref{F4} for the same value $F_0=-1.05$ that, hopefully,  minimizes the higher-order
corrections even for a timelike momentum. Among the main features of the spectral function we observe
the lack of any sharp peak or pole and the violation of positivity. Fig.\ref{F4} is in qualitative agreement
with the numerical analytic continuation of Ref.\cite{dudal14}, showing the same pattern of a negative
minimum followed by a positive maximum and a negative large-energy part that reaches a minimum and then
increases again asymptotically. However, the maximum is more peaked in Fig.\ref{F4} and is shifted at a larger
value  $p^2\approx 0.54$ GeV$\>^2\approx m^2$ compared to $p^2\approx 0.2$ GeV$\>^2$ of Ref.\cite{dudal14}.
A very sharp peak at $p^2\approx 0.4$ GeV$\>^2$ was predicted in Ref.\cite{straussDSE}
by a numerical calculation based on DSE formalism in Minkowski space. In that work the violation of
positivity was found to set in only after the peak for $p^2>0.4$ GeV$\>^2$, while in Fig.\ref{F4} a negative range 
also appears for small $p^2$, in agreement with Ref.\cite{dudal14}.

Out of the real axis, in the complex plane, the propagator has two conjugated poles at 
$(\Re p^2,\Im p^2)\approx(0.16,\pm 0.60)$ GeV$^2$, close to the imaginary axis, as predicted
by the {\it i-particle} scenario\cite{iparticle} emerging from the refined version\cite{dudal08,dudal08b,dudal11}
of the Gribov-Zwanziger model\cite{GZ}. It is remarkable that, by a fit of the data, in Ref.\cite{sorella10}
a complex mass was extracted from that model, $m^2=(0.1685,\pm 0.4812)$ GeV$^2$, not too far from
the present result. We must mention that no complex poles were found by the numerical study of Ref.\cite{straussDSE}.

\begin{figure}[t] 
\centering
\includegraphics[width=0.35\textwidth,angle=-90]{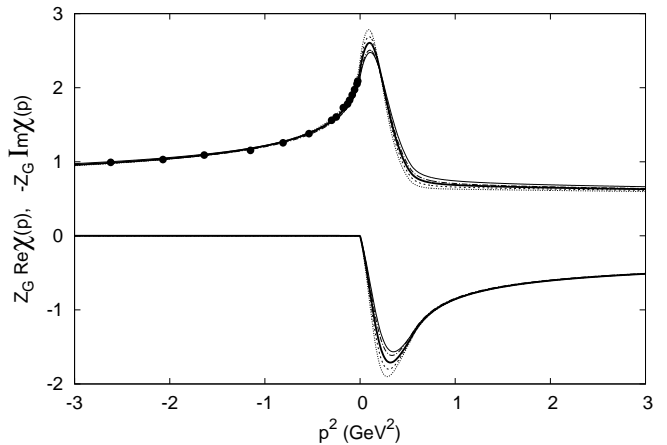}
\caption{Real part $\Re\chi$ and imaginary part $-\Im\chi=\pi p^2\rho$ of the ghost dressing function
according to Eq.(\ref{dress3}) for $m=0.73$ GeV and
several values of $G_0$ in the range $0.2<G_0<0.3$. The points are the lattice data 
of Ref.\cite{bogolubsky} ($N=3$, $\beta=5.7$, $L=80$). The best agreement with the data points is 
obtained for $G_0=0.24$ (solid line). The dressing function is scaled by a finite
renormalization constant $Z_G$.}
\label{F5}
\end{figure}

Another interesting feature of Fig.\ref{F4} is the existence of a cusp at the two-particle threshold 
$p^2=(2m)^2\approx (1.46)^2$ GeV$\>^2$. That is roughly the point where the real part turns negative.
Overall, the real part of the gluon propagator in Fig.\ref{F2} is in qualitative agreement with the phenomenological
propagator in Fig.11 of Ref.\cite{shakinPRD}, based on the existence of a BRST-invariant condensate of 
dimension two\cite{kondo} that reduces to $A_{min}^2$ in the Landau gauge. 
In agreement with Ref.\cite{shakinPRD}, there is not a pole but just a point where $\Re \Delta=0$. Then the
real part turns negative, reaches a minimum and increases asymptotically towards zero.

The one-loop ghost propagator, by Eq.(\ref{dress3}) maintains a pole at $p^2=0$. The analytic
continuation $s=-p^2/m^2-i\varepsilon$  yields
\begin{align}
\Re {\cal G}(p^2+i\varepsilon)&=\frac{\Re \chi (p^2)}{p^2}\nn\\
\Im {\cal G}(p^2+i\varepsilon)&=\frac{\Im \chi (p^2)}{p^2}-\pi\chi(0)\>\delta(p^2)
\label{RIG}
\end{align}
and we can define a spectral function on the cut
\BE
\rho(p^2)=-\frac{1}{\pi}\Im {\cal G}(p^2+i\varepsilon)=
\chi(0)\>\delta(p^2)-\frac{1}{\pi}\frac{\Im \chi(p^2)}{p^2}
\label{rho2}
\EE
which has a continuous term given by the imaginary part of the dressing function divided
by $-p^2$.  The details of the continuous term of the spectral function are shown in Fig.\ref{F5}
by the direct plot of $-\Im \chi$ , together with the real part $\Re \chi$ and the lattice
data of Ref.\cite{bogolubsky} ($N=3$, $\beta=5.7$, $L=80$). We observe that the discrete and the continuous
terms have the opposite sign in Eq.(\ref{rho2}), violating the positivity condition. In the Euclidean range
$p^2<0$, the ghost dressing function is not very sensitive to a change of the additive constant $G_0$. 
In Fig.\ref{F5}, a change of $G_0$ in the range $0.2<G_0<0.3$ is compensated by a change of the finite renormalization 
constant $Z_G$, so that $Z_G\chi(p^2)$ stays on the lattice data points. That is probably a consequence of
being massless. The best agreement is found for $G_0=0.24$ and is shown as a solid line in Fig.\ref{F5}. 
The imaginary
part has a wide peak at $p^2\approx (0.56)^2$ GeV$\>^2$ and never changes sign.

\section{Chiral QCD}

The inclusion of a set of chiral quarks requires the calculation of the quark loops contributing to
the gluon polarization and the quark self-energy $\Sigma_q$. 
At one-loop, we must add all the graphs of Fig.\ref{F1}.

\subsection{Ghost Propagator}

The ghost sector of QCD provides a stringent test for the massive expansion. As shown in Fig.\ref{F1},
there are no one-loop graphs with quark lines that contribute to the ghost self-energy $\Sigma_{gh}$.
Thus the one-loop ghost dressing function of QCD is the same as that of pure Yang-Mills theory
and is given by Eq.(\ref{dress3}). The weight of higher order loops can be estimated by comparing
the data of lattice simulations in the presence of quenched and unquenched quarks, as the difference can 
only arise from terms that are neglected in the expansion.

\begin{figure}[t] 
\centering
\includegraphics[width=0.35\textwidth,angle=-90]{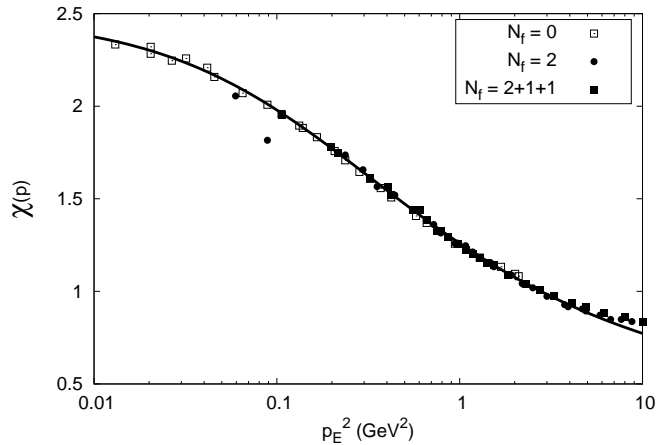}
\caption{The ghost dressing function $\chi(s)$ of Eq.(\ref{dress3}) is shown as a function of the 
Euclidean momentum $p_E^2=s\> m^2$ for $m=0.7$ GeV and $G_0=0.195$.
The points are lattice data extracted from Fig.1 of Ref.\cite{binosi12} and renormalized by
the factors $Z_G=0.86$ ($N_f=0$, open squares, originally taken from Ref.\cite{bogolubsky}), 
$Z_G=0.78$ ($N_f=2$, two light quarks, circles),
$Z_G=0.78$ ($N_f=2+1+1$, two light and two heavy quarks, filled squares). The energy units of the
quenched data ($N_f=0$) have been multiplied by a factor $1.12$ with respect to the original data
of Ref.\cite{bogolubsky} (a point at $p=1$ GeV in the figure was at $p=1.12$ GeV in the original figure).}
\label{A1}
\end{figure}

Actually, that difference is very small and not very sensitive to number of quarks and their mass, 
so that we can compare the unquenched data of Ref.\cite{binosi12} for two light quarks with the quenched data of
Ref.\cite{bogolubsky}, as already done in Ref.\cite{binosi12}. It is not even obvious that
there is a change at all, as the only difference seems to be a scale factor $\approx 1.12$ 
in the energy units of the two data sets. In Fig.\ref{A1} we show that, by rescaling the units of the quenched
set of Ref.\cite{bogolubsky} ($N_f=0$) and renormalizing the dressing function by the factor $Z_G=0.86$
for $N_f=0$ and $Z_G=0.78$ for $N_f=2$, all the data in Fig.3 of Ref.\cite{binosi12} collapse on the
same curve given by Eq.(\ref{dress3}) for $m=0.7$ GeV and $G_0=0.195$. Once more, the ghost dressing function
seems to be described by the universal function $G(s)$ of Eq.(\ref{FGx}) irrespective of coupling, renormalization
and even quark number. In other words, the inverse dressing function is entirely determined up to an
additive constant and a renormalization factor. The extension to full QCD of the same scaling property 
that was first predicted\cite{ptqcd,ptqcd2} for pure Yang-Mills theory is an indirect proof that higher 
order loops are not very relevant in the ghost sector.

Having compared data sets of different authors, we cannot exclude that a slight accidental difference 
has occurred in the choice of the energy units. However, the factor $1.12$, which is required for 
rescaling the quenched data over the 
unquenched plot in Fig.\ref{A1}, seems to be too large for being accidental. 
Then, we assume that it is a genuine rescaling produced by the quark loops. 
In that case, the same
curve in the figure can be seen as the plot of the one-loop dressing function in Eq.(\ref{dress3}) 
with $m=0.7$ GeV for $N_f=2$ (or larger) and with $m=1.12\times 0.7=0.78$ GeV for $N_f=0$.
If we believe that a judicious choice of $m$ minimizes the effects of higher loops, then we conclude
that going from the quenched to the unquenched theory the optimal value of $m$ decreases a little in
order to take into account the neglected quark loops. Thus, even without the inclusion of new graphs, the
one-loop calculation can feel the effect of quarks by a very small shift of the mass parameter.

Having added no new graphs to the self-energy, the analytic properties of the ghost propagator are the same of 
pure Yang-Mills theory, as discussed in the previous section.

\subsection{Gluon Propagator}

At one-loop, the gluon polarization of the full theory is obtained from the result for pure Yang-Mills
theory by just adding the quark loops of Fig.\ref{F1}.

The second-order uncrossed quark loop reads
\BE
\Pi_{ab}^{\mu\nu}(p)=-\delta_{ab}\frac{g^2}{2}\Tr\int\qqqi \gamma^\mu S_M(q) \gamma^\nu S_M (p+q).
\EE
Dropping the color indices, this term is just $g^2/(2e^2)$ times the ordinary electron loop of QED
and can be evaluated by dimensional regularization. Setting $d=4-\epsilon$ the result is transverse
and, summing over $N_f$ quarks, it can be written as
\BE
\Pi_{ab}^{\mu\nu}(p)=\delta_{ab} t^{\mu\nu}(p) \Pi (p)=\delta_{ab} t^{\mu\nu}(p) 
\left[\Pi^\epsilon (p)+\Pi^f (p)\right]
\EE
where the diverging part reads
\BE
\Pi^\epsilon(p)=\frac{N_f \alpha_s}{3\pi} p^2 \left(\frac{1}{\epsilon}+\log\frac{\mu}{M}\right)
\label{1leps}
\EE
and the finite part is
\BE
\Pi^f(p)=\frac{N_f \alpha_s}{3 \pi} p^2\ f_0(-p^2/M^2)
\label{1lfin}
\EE
in terms of the adimensional function
\BE
f_0(s)=\left(\frac{s-2}{2s}\right) L(s)-\frac{2}{s}
\label{f0}
\EE
and of the logarithmic function
\BE
L(s)=\sqrt{\frac{4+s}{s}}\log\frac{\sqrt{4+s}-\sqrt{s}}{\sqrt{4+s}+\sqrt{s}}.
\label{L}
\EE
In the definition of $f_0$, Eq.(\ref{f0}), we dropped a scheme-dependent additive constant that
would depend on the definition of the arbitrary scale $\mu$ in Eq.(\ref{1leps}) and can be absorbed by 
the finite part of the wave-function renormalization constant.
In the $\overline{MS}$ scheme, denoting by $\delta Z=Z-1$ the one-loop contribution to a
renormalization constant, the divergence in the quark loop $\Pi^\epsilon$ is canceled by
a term
\BE
\delta Z_A=-\frac{N_f \alpha_s}{3\pi\epsilon}=-\frac{N_f g^2}{12\pi^2}\frac{1}{\epsilon}
\label{ZA}
\EE
to be added to $Z_A$ in Eq.(\ref{Z}).
As a check we observe that in the Landau gauge the ghost-gluon vertex is regular\cite{taylor}
and the renormalization constant $Z_g$ of the coupling can be extracted from the wave-function
renormalization of the two-point functions, yielding by Eq.(\ref{Z}) and Eq.(\ref{ZA}) for $N=3$
\BE
\delta Z_g=\frac{1}{2}\left[\delta Z_A+2\delta Z_\omega\right]=
\frac{g^2}{16\pi^2} \frac{1}{\epsilon}\left[11-\frac{2N_f}{3}\right]
\label{Zg}
\EE
which is the well known standard result for QCD.

Adding the crossed quark-loop in Fig.\ref{F1}, the sum of all quark loops follows 
by Eq.(\ref{crossedsum}) for $n=1$
\BE
\Pi_{tot}=\left[1-M\frac{\partial}{\partial M}\right] \Pi.
\label{Pitot}
\EE
As before we can write $\Pi_{tot}=\Pi^\epsilon_{tot}+\Pi^f_{tot}$
where the diverging part $\Pi^\epsilon_{tot}=\Pi^\epsilon$ does not change and is still given by
Eq.(\ref{1leps}) because the  term $1/\epsilon$ does not depend on $M$.
Thus, the divergence is canceled by the same standard wave function renormalization constant $\delta Z_A$
of Eq.(\ref{ZA}). Then, Eq.(\ref{Zg}) still holds and the expansion has the same behavior as the standard 
perturbative expansion in the UV.

The finite part follows by a simple derivative and can be written as
\BE
\Pi^f_{tot}=\frac{N_f \alpha_s}{3\pi} p^2\ f(-p^2/M^2)
\label{1ltot}
\EE
in terms of the adimensional function
\BE
f(s)=f_0(s)+2s\left(\frac{\partial f_0}{\partial s}\right)
\EE
that has the explicit expression
\BE
f(s)=
\frac{4}{s}+\left(\frac{s^2+2s+16}{2s(4+s)}\right) L(s)
\label{f1}
\EE
up to an irrelevant additive constant that is absorbed in the finite wave-function
renormalization.

In the limit $s\to 0$ the function $L(s)\approx -2-s/6$ and $\Pi^f_{tot}\approx p^2(N_f\alpha_s)/(18\pi)
\to 0$ so that the quark loop does not contribute to the gluon mass. On the same footing, we see that the 
photon does not acquire any mass term from the fermions and since there are no photon loops in QED, the
massive expansion does not give any mass to the photon.

As for pure Yang-Mills theory, the gluon dressing function is still given by Eq.(\ref{dress3})
provided that  a quark correction $\Delta F(s)$ is added to the function $F(s)$
\BE
\Delta F (s)=\frac{4N_f}{9N}\>f(m^2s/M^2)
\label{DF}
\EE
where $N=3$ for QCD.

\begin{figure}[t] 
\centering
\includegraphics[width=0.35\textwidth,angle=-90]{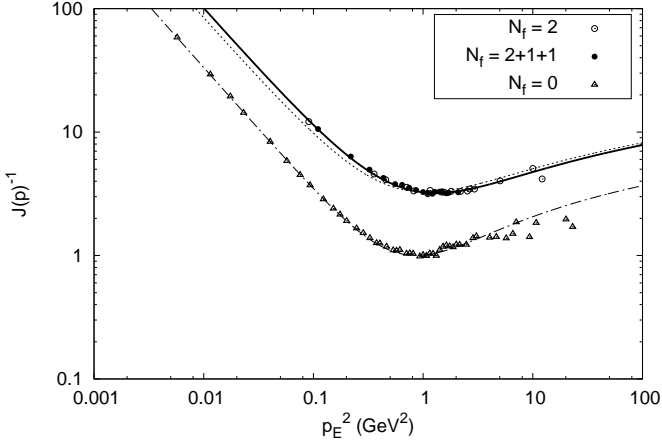}
\caption{The inverse gluon dressing function $1/J(s)=F(s)+\Delta F(s)+F_0$, as arising from 
Eqs.(\ref{dress3}),(\ref{DF}), is plotted as a function of the Euclidean momentum $p_E^2=s\> m^2$
for $N_f=2$, $m=0.8$ GeV, $M=0.65$ GeV (optimal set, solid line), for $N_f=2$, $m=0.73$ GeV, $M=0.65$ GeV 
(sub-optimal, dotted line) and for $N_f=0$, $m=0.73$ GeV (pure Yang-Mills, $\Delta F=0$,  dashed line).
The constant $F_0$ is $-0.65$ for $N_f=2$ and $-1.05$ for $N_f=0$ \cite{ptqcd2}.
The circles are lattice data extracted from Fig.1 of Ref.\cite{binosi12} and scaled as described in the
text. The triangles are lattice data extracted from Fig.2 of Ref.\cite{bogolubsky}.}
\label{B1}
\end{figure}

According to our strategy, the expansion has to be optimized by a comparison with the data of 
lattice simulations in the Euclidean space.
First of all, let us explore the scaling properties of the inverse dressing function $J^{-1}$
that arise from Eq.(\ref{dress3}) when modified by Eq.(\ref{DF}), now including the quark loops.
The inverse gluon dressing function is shown in Fig.\ref{B1} together with the result found
for pure Yang-Mills theory ($\Delta F=0$) in Refs.\cite{ptqcd,ptqcd2}. The pronounced minimum
is useful for pinpointing the energy scale by a comparison with other data.

We expect that,
up to an additive constant and an irrelevant renormalization factor, Eq.(\ref{dress3}) should be quite 
general and predict a scaling behavior satisfied by all data of lattice simulations,
even in the presence of small bare masses for the quarks. The effect of small bare
masses is negligible compared with the large dynamical masses in the infrared, so that no dramatic difference
is expected between chiral and light quarks. On the other hand, the effects of heavy quarks are going
to be suppressed in the quark loops. Thus, having no lattice data for the gluon propagator in the chiral limit,
we are comparing Eq.(\ref{DF}) with the simulations of Ref.\cite{binosi12} for the case of two light quarks. 

In Fig.{\ref{B1}} the data points are extracted from Fig.1 of Ref.\cite{binosi12} and rescaled by different 
renormalization factors and additive constants. In more detail, the data for $N_f=2$ (two light quarks) have been
divided by $0.18$, while the data for $N_f=2+1+1$ (two light quarks and two heavy quarks) are divided by 
$0.24$ and increased by adding the constant $0.45$, which is equivalent to a different choice for the constant
$F_0$ in Eq.(\ref{dress3}). It is remarkable that the data collapse on the curve predicted by 
Eqs.(\ref{dress3}),(\ref{DF}) with $N_f=2$ (two chiral quarks) and $m=0.8$ GeV. The plot is not very sensitive to
the choice of the mass $M$ and a slight change of $M$ is compensated by a minor change of the constant $F_0$
in Eq.(\ref{dress3}). In Fig.\ref{B1} we are using $F_0=-0.65$ and the mass $M=0.65$ GeV that will turn out to be a 
reasonable choice for describing the quark propagator in the next section.
More than the mass $M$, it is the number of light quarks $N_f$ that has an effect on the position of the
minimum in the figure. Compared to pure Yang-Mills theory ($N_f=0$) we observe a $25\%$ shift 
of the minimum towards higher energies for $N_f=2$. No further shift is observed for the inclusion of the 
heavier quarks ($N_f=2+1+1$).
We find that a $14\%$ shift is given by the extra term $\Delta F$ in Eq.(\ref{DF}), coming from the
quark loops, while the residual $10\%$ shift in Fig.\ref{B1} is obtained by increasing the mass scale up to
$m=0.8$ GeV, compared to the value $m=0.73$ GeV that was required for $N_f=0$.
Assuming that the agreement with the data means that higher-loop corrections are negligible, we would extract
an optimal value $m=0.8$ GeV, while the best choice for the other scale $M$ remains almost undetermined 
because of the small sensitivity to that mass. The slight increase of $m$, going from $N_f=0$ to $N_f=2$,
is in the opposite direction of the small shift that was required for the best description of the
ghost dressing function in the previous section. That is not surprising, as there is no reason
why the effects of higher loops should be minimal for all the propagators at the same value of $m$. While that
would be desirable, those effects could have a different sign and require a compromise. For future reference, as
an average choice, we could just leave $m$ unchanged and take the value $m=0.73$ GeV that was optimal for $N_f=0$.
As shown by the dotted line in Fig.\ref{B1} the agreement with the data is still acceptable (as it is for
the ghost dressing function).

\begin{figure}[t] 
\centering
\includegraphics[width=0.35\textwidth,angle=-90]{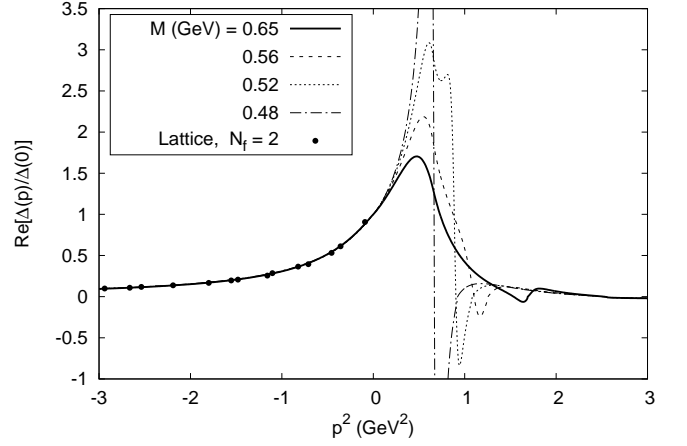}
\caption{The real part of the gluon propagator is evaluated by setting $s=-p^2/m^2-i\varepsilon$
in Eqs.(\ref{dress3}),(\ref{DF}),
for $m=0.80$ GeV and several values of $M=0.48,0.52,0.56,0.65$ GeV. 
The constant $F_0$ varies in
the range $-0.65<F_0<-0.6$ in order to keep all curves on the lattice data in the Euclidean
space, for $p^2<0$. The data points are extracted from Fig.1 of Ref.\cite{binosi12} for $N_f=2$.
The propagator is normalized by its finite value at $p^2=0$.}
\label{B2}
\end{figure}

\begin{figure}[t] 
\centering
\includegraphics[width=0.35\textwidth,angle=-90]{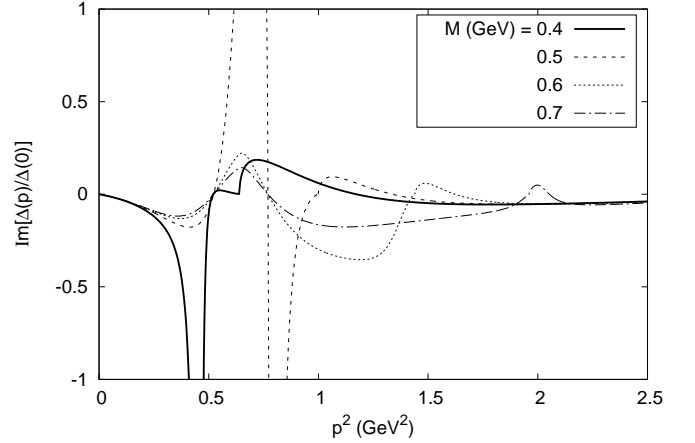}
\caption{The imaginary part of the gluon propagator is evaluated by setting $s=-p^2/m^2-i\varepsilon$
in Eqs.(\ref{dress3}),(\ref{DF}), for $m=0.80$ GeV, $F_0=-0.6$ and several values of $M=0.4,0.5,0.6,0.7$ GeV. 
The propagator is normalized by its finite value at $p^2=0$.}
\label{B3}
\end{figure}

\begin{figure}[t] 
\centering
\includegraphics[width=0.35\textwidth,angle=-90]{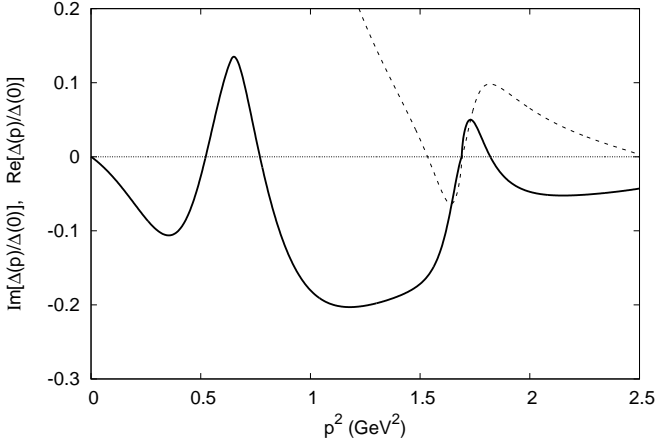}
\caption{The imaginary part of the propagator is evaluated by setting $s=-p^2/m^2-i\varepsilon$
in Eqs.(\ref{dress3}),(\ref{DF}), for the optimal set $m=0.80$ GeV, $M=0.65$ GeV,  $F_0=-0.65$ (solid line).
The dashed line is a detail of the real part. The propagator is normalized by its finite value at $p^2=0$.}
\label{B4}
\end{figure}

The real part of the gluon propagator is shown in Fig.\ref{B2} for $s=-p^2/m^2-i\varepsilon$. While rather
insensitive to the choice of $M$ in the Euclidean space, the shape of the propagator depends 
on $M$ when plotted as a function of the time-like momentum $p^2>0$. 
The data points in the figure are the same lattice data of Ref.\cite{binosi12} for $N_f=2$ that were
already used in Fig.\ref{B1}. 
We observe the presence of a 
positive peak at $p^2\approx m^2$ and a negative peak just before the two-particle threshold $p^2\approx (2M)^2$
where the real part of the propagator changes sign and becomes positive (see also the details in Fig.\ref{B4}).
When $M$ decreases the negative peak comes closer to the positive one and they merge eventually, when
$2M\approx m$. Below that point, for $M<0.5$ GeV they give rise to a sharp peak.
That peak cannot be physical, because its spectral weight is negative, as shown in Fig.\ref{B3} where the
imaginary part of the propagator is displayed. It defines a spectral density according to Eq.(\ref{rho}) and
its positivity violation is what we expected for a confined degree of freedom. At variance with pure
Yang-Mills theory, we observe the presence of a two-particle threshold at $p^2\approx (2M)^2$ where
the spectral function turns positive for a while. Some details of the spectral function are shown
in Fig.\ref{B4} for the optimal set $m=0.8$ GeV, $M=0.65$ GeV of Fig.\ref{B1}, to be compared to
Fig.\ref{F4} for pure Yang-Mills theory.

Besides being more rich on the real positive axis $p^2>0$, for $N_f=2$ the unquenched gluon propagator has
more poles in the complex plane. For the optimal set $m=0.8$ GeV, $M=0.65$ GeV We find two pairs of 
conjugated poles at $p^2\approx (1.69,\pm 0.1)$ GeV$^2$ and $p^2\approx (0.54,\pm 0.52)$ GeV$^2$. 
Changing the mass $M$, the first pair, closer to the real axis, moves according to $\Re p^2\approx (2M)^2$, 
while the second pair stays close to $\Re p^2\approx m^2$.

\subsection{Quark Propagator}

At one loop and third order, the quark self energy $\Sigma_q$ is given by the tree term 
$\delta \Gamma_q=-M$ and 
three one-loop graphs, as shown in Fig.\ref{F1}.
The sum of the one-loop graphs
follows from Eq.(\ref{crossedsum}) 
\BE
\Sigma (p)=
\left[1-m^2\frac{\partial}{\partial m^2} -M \frac{\partial}{\partial M}\right]
\>\Sigma^{(1)} (p)
\label{cross2}
\EE
where $\Sigma^{(1)} (p)$ is the standard (uncrossed) one-loop graph and the derivatives
give the crossed graphs. On general grounds, we can write all self-energy graphs in terms of
Lorentz scalar functions
\begin{align}
\Sigma^{(1)} (p)&=\Sigma_M^{(1)}(p)+\fsl{p}\>\Sigma_p^{(1)}(p)\nn\\
\Sigma (p)&=\Sigma_M (p)+\fsl{p}\>\Sigma_p(p).
\label{scalar}
\end{align}

In the dressed quark propagator $S(p)$
the mass $M$ is canceled by the tree term $\delta \Gamma_q=-M$,
yielding by Eq.(\ref{Deltam})
\BE
S(p)^{-1}=S_M(p)^{-1}-\delta\Gamma_q-\Sigma (p)=\fsl{p}-\Sigma (p).
\label{Sdress}
\EE
By insertion of Eq.(\ref{scalar}) the dressed quark propagator reads
\BE
S(p)=\frac{Z_q}{\fsl{p}\left[1-\Sigma_p(p)\right]-\Sigma_M (p)}
\label{Sdress2}
\EE
where $Z_q$ is a finite wave-function renormalization constant.
At one-loop, we can set
\BE
Z_q=1+\delta Z_q\approx \frac{1}{1-\delta Z_q}
\label{Zq}
\EE
and defining a subtraction point $\mu$ according to
\BE
\delta Z_q=-\Sigma_p (\mu)
\label{mudef}
\EE
we can write 
\BE
S(p)^{-1}=\fsl{p}\left[1-\left(\Sigma_p(p)-\Sigma_p(\mu)\right)\right]-\Sigma_M (p).
\label{Sdress3}
\EE
On the other hand, by the usual parametrization in terms of the scalar functions $Z(p)$, $M(p)$,
the dressed propagator reads
\BE
S(p)=\frac{Z(p)}{\fsl{p}-M(p)}
\label{Sscalar}
\EE
and by comparison with Eq.(\ref{Sdress3})we can write
\begin{align}
Z(p)^{-1}&=1-\left[\Sigma_p(p)-\Sigma_p(\mu)\right]+{\cal O} (\alpha_s^2)\nn\\
M(p)&=\frac{\Sigma_M (p)}{\left[1-\left(\Sigma_p(p)-\Sigma_p(\mu)\right)\right]}+{\cal O} (\alpha_s^2).
\label{ZM}
\end{align}
We observe that, being the ratio of the two scalar functions, the exact mass function $M(p)$ does not
depend on the choice of the renormalization constant. Thus, a slight dependence on the subtraction
point $\mu$ would be a natural consequence of the one-loop approximation. We anticipate that, at one-loop, 
the function $\Sigma_p(p)$ is almost constant in the UV, then $M(p)$ in Eq.(\ref{ZM}) does not depend on
the subtraction point provided that we take $\mu$ large enough. We will safely take $\mu\to \infty$ so that
$S(p)^{-1}\to \fsl{p}$ in the UV.

By its definition in Eq.(\ref{Sscalar}), the exact scalar function $Z(p)$ depends on the
wave-function renormalization. According to Eq.(\ref{ZM}), the subtraction point $\mu$ is defined by 
$Z(\mu)=1$.

Some details on the calculation of the explicit functions $\Sigma_M$, $\Sigma_p$ are given in the Appendix.
The standard uncrossed one-loop graph, the first in Fig.\ref{F1}, can be evaluated analytically by dimensional
regularization and the resulting scalar functions $\Sigma_M^{(1)}$, $\Sigma_p^{(1)}$ are reported 
in Appendix B. At variance with the standard perturbative expansion, a mass is included in the
gluon propagator $\Delta_m$ inside the loop, according to Eq.(\ref{Deltam}).
The results coincide with that reported by other authors\cite{tissier14}.

The function $\Sigma_p^{(1)} (p)$ is finite and vanishes in the limit $m\to 0$. Even for $m\not=0$ the
standard one-loop contribution to $Z(p)$ is very small and shows a wrong decreasing
behavior as a function of the Euclidean momentum $s=-p^2/m^2$ in comparison with the lattice data.
That behavior was shown to be a consequence of two-loop corrections\cite{tissier14} 
that are not negligible because of the very tiny one-loop contribution.

Including all the crossed graphs in Fig.\ref{F1}, 
the full scalar function $\Sigma_p (p)$ is given by Eq.(\ref{cross2})
\BE
\Sigma_p (p)=
\left[1-m^2\frac{\partial}{\partial m^2} -M \frac{\partial}{\partial M}\right]
\>\Sigma_p^{(1)} (p).
\label{cross3}
\EE
The result is not very illuminating and the explicit analytical expression is derived in  Appendix C.

As shown in Eq.(\ref{sigma1}), the other scalar (uncrossed) one-loop function 
$\Sigma_M^{(1)}$ has a diverging part
\BE
\Sigma_M^{(1)}\sim \frac{\alpha_s}{\pi}M\left(\frac{2}{\epsilon}+\log\frac{\mu^2}{m^2}\right)
\label{diverg}
\EE
that usually requires a mass counterterm for its cancellation. In the chiral limit, we have no mass
counterterm for the quarks. However, the divergence is canceled by the
crossed loops in the total self-energy according to Eq.(\ref{cross2})
\BE
\Sigma_M (p)=
\left[1-m^2\frac{\partial}{\partial m^2} -M \frac{\partial}{\partial M}\right]
\>\Sigma_M^{(1)} (p).
\label{cross4}
\EE

\begin{figure}[t] 
\centering
\includegraphics[width=0.35\textwidth,angle=-90]{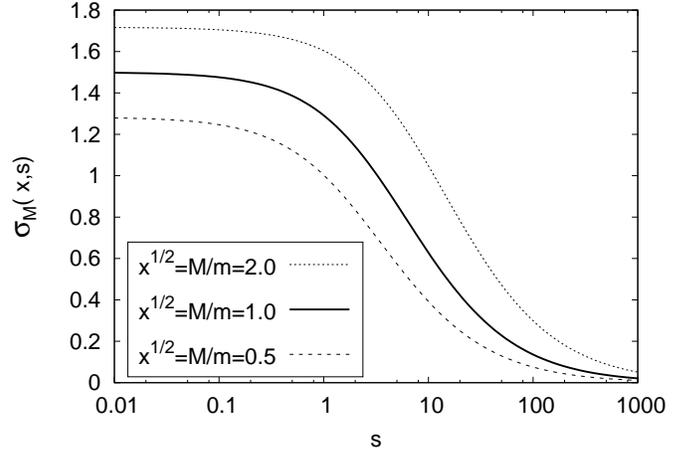}
\caption{The adimensional scalar function $\sigma_M (x,s)$ of Eq.(\ref{sigmatot}) is shown as a function of the 
Euclidean momentum $s=-p^2/m^2$ for different values of the mass ratio $x=M^2/m^2$ corresponding to $M=2m$ (top),
$M=m$ (center) and $M=m/2$ (bottom).}
\label{C1}
\end{figure}

\begin{figure}[t] 
\centering
\includegraphics[width=0.35\textwidth,angle=-90]{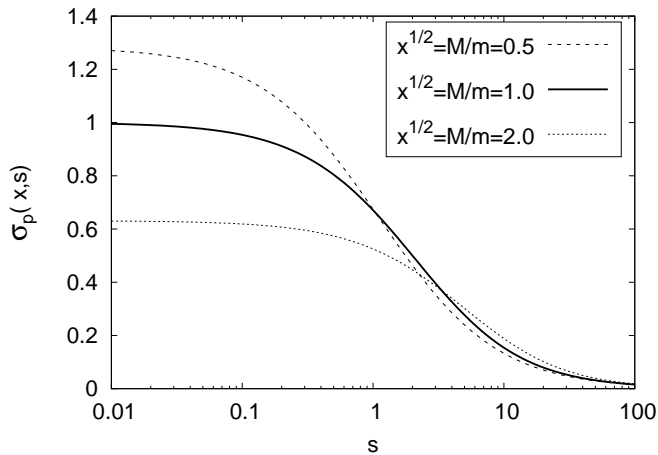}
\caption{The adimensional scalar function $\sigma_p (x,s)$ of Eq.(\ref{sigmatot}) is shown as a function of the 
Euclidean momentum $s=-p^2/m^2$ for different values of the mass ratio $x=M^2/m^2$ corresponding to $M=2m$ 
(dotted line), $M=m$ (solid line) and $M=m/2$ (dashed line).}
\label{C2}
\end{figure}

Inserting Eq.(\ref{diverg}) into Eq.(\ref{cross4}) we see that the diverging part cancels exactly,
yielding a finite function $\Sigma_M(p)$
without the need for any mass counterterm that would spoil the chiral symmetry of the original Lagrangian.
The explicit analytical result is derived in Appendix C and can be written in terms of functions of the Euclidean
momentum $s=-p^2/m^2$ and the squared mass $x=M^2/m^2$ 
\begin{align}
\Sigma_M (p)&=\left(\frac{\alpha_s}{\pi}\right) M \>\sigma_M (x,s)\nn\\
\Sigma_p (p)&=\left(\frac{\alpha_s}{3 \pi}\right) \>\sigma_p \>(x,s)
\label{sigmadef}
\end{align}
where the adimensional functions $\sigma_p$, $\sigma_M$ include all the loops in Fig.\ref{F1}, are
finite and do not depend on any parameter. Their explicit expressions are given in Eqs.(\ref{sigmatot})
where the zero of $\sigma_p$ is set by $\sigma_p(x,\infty)=0$. Their plot is shown in Fig.\ref{C1} and
Fig.\ref{C2} for a few values of the mass ratio $x$.

\begin{figure}[t] 
\centering
\includegraphics[width=0.35\textwidth,angle=-90]{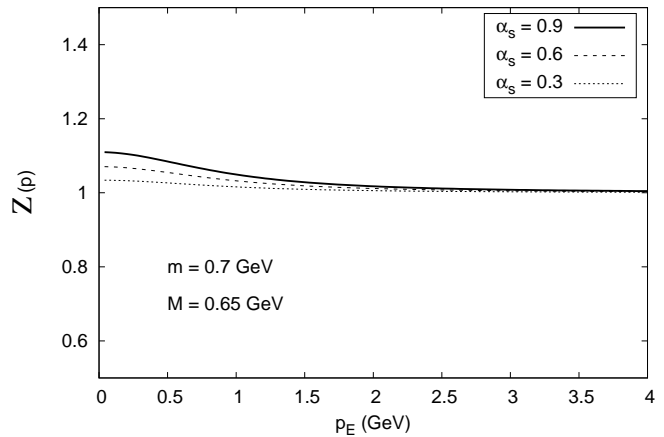}
\caption{The scalar function $Z(p)$ as a function of the Euclidean momentum $p_E$ for three different values 
of the strong coupling $\alpha_s=0.3$, $0.6$, $0.9$.} 
\label{C3}
\end{figure}

In terms of these functions, taking $\mu\to\infty$, Eq.(\ref{ZM}) can be recast as
\begin{align}
Z(p)&=\frac{\sigma_0}{\sigma_0-\sigma_p(x,s)}\nn\\
M(p)&=3M \left[\frac{\sigma_M (x,s)}{\sigma_0-\sigma_p(x,s)}\right]
\label{ZM2}
\end{align}
where the inverse coupling $\sigma_0=3\pi/\alpha_s$ is the only parameter besides the mass scales $m$, $M$.
Even if the functions $\sigma_p$, $\sigma_M$ are of the same order, the effect of $\sigma_p$ on the function
$Z(p)$ is very small, giving a flat, almost constant {\it decreasing} function, as shown in Fig.\ref{C3}. Thus 
the mass function $M(p)$ is not sensitive to the subtraction point and is basically 
determined by the function $\sigma_M$. As discussed in Ref.\cite{tissier14}, the wrong decreasing
behavior of $Z(p)\approx 1$, when compared with the outcome of lattice simulations\cite{bowman05}, 
means that the effect of two-loop corrections might overcome the tiny effect of one-loop terms on the
normalization of the propagator.

At variance with gluon and ghost propagators, that have no explicit dependence on the
coupling because of their multiplicative renormalization, the mass function $M(p)$ does not
depend on the normalization of the quark propagator and has an explicit dependence on 
the strong coupling $\alpha_s$ in Eq.(\ref{ZM2}). Here is where the limits of a fixed-coupling expansion
emerge, since a running of the coupling is mandatory for connecting the low energy phenomenology
with the high-energy values of the coupling. Thus it is remarkable that, by using reasonable values
for the coupling below 1 GeV, $\alpha_s\approx 0.5-1.0$\cite{bogolubsky,twoloop}, the chiral Lagrangian
develops a dynamical mass for the quarks, with the correct qualitative behaviour and magnitude of
the constituent-quark masses. We stress once more that the quark mass does not arise trivially by the insertion
of a mass in the undressed propagator, since that mass is canceled at tree-level in Eq.(\ref{Sdress}). Thus the 
mass arises from the loops. While an overall energy scale can only be fixed by the phenomenology, the ratio
$M/m$ can be regarded as a variational parameter: something the exact theory should not depend on, to be fixed by
some optimization strategy. In this paper we pursue the strategy of optimizing the mass scales by a
comparison with the lattice data, in order to predict the behavior of the propagators in Minkowski space
by analytic continuation.
The data of lattice simulations seems to favor a ratio $M/m$ not too far from one,
which suggests a link between the dynamical generation of the gluon mass and the dynamical breaking
of the chiral symmetry.

\begin{figure}[t] 
\centering
\includegraphics[width=0.35\textwidth,angle=-90]{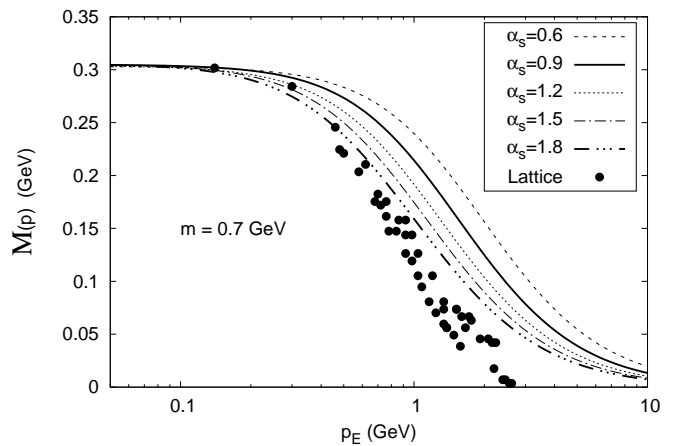}
\caption{The mass function $M(p)$ as a function of the Euclidean momentum $p_E$ for several values 
of the coupling
$\alpha_s=0.6,0.9,1.2,1.5,1.8$ and $m=0.7$ GeV. The points are lattice data for unquenched chiral quarks and have been 
extracted from Fig.4 of Ref.\cite{bowman05}. For each coupling the mass parameter $M$ is fixed by requiring that
$M(0)\approx 0.32$ GeV, yielding $M=0.94, 0.65,0.49,0.39,0.318$ GeV, respectively.} 
\label{C4}
\end{figure}

\begin{figure}[t] 
\centering
\includegraphics[width=0.35\textwidth,angle=-90]{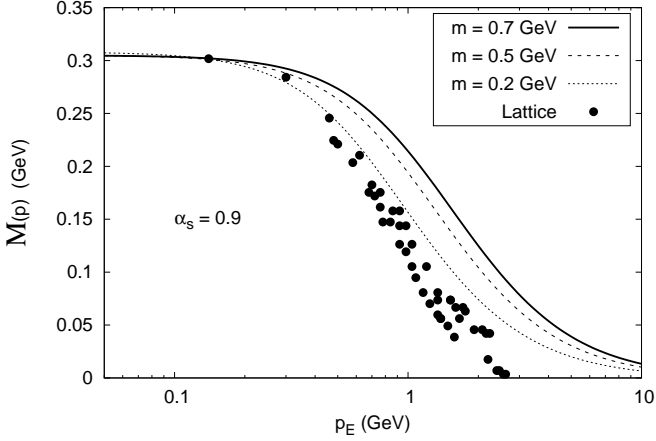}
\caption{The mass function $M(p)$ as a function of the Euclidean momentum $p_E$ for three values of the 
mass scale $m=0.7,0.5,0.2$ GeV and $\alpha_s=0.9$. The mass parameter $M$ takes the values
$M=0.65,0.62,0.57$ GeV, respectively. The points are the same lattice data of Fig.\ref{C4}.}
\label{C5}
\end{figure}

\begin{figure}[t] 
\centering
\includegraphics[width=0.35\textwidth,angle=-90]{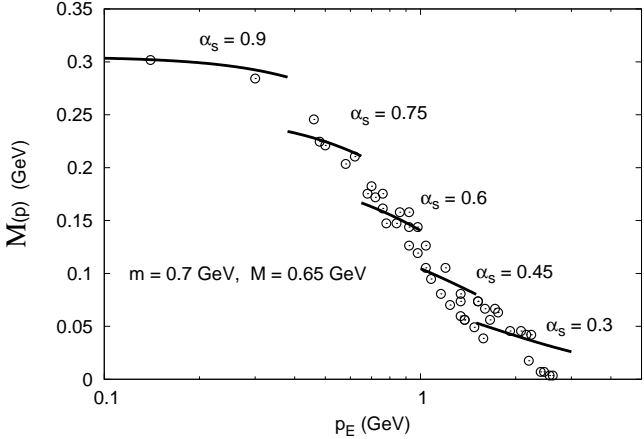}
\caption{The mass function $M(p)$ as a function of the Euclidean momentum $p_E$ for 
$m=0.7$ GeV and $M=0.65$ GeV. The coupling $\alpha_s$ is taken as a piecewise constant function, 
decreasing from $\alpha_s=0.9$ at $p_E<0.4$ GeV down to $\alpha_s=0.3$ at $p_E\approx 2$ GeV. 
The points are the same lattice data of Fig.\ref{C4}.}
\label{C6}
\end{figure}

The mass function $M(p)$ of Eq.(\ref{ZM2}) is displayed in Fig.\ref{C4} as a function of the Euclidean
momentum for several values of the coupling. For each coupling, the mass parameter $M$ has been fixed by
requiring that $M(0)\approx 0.32$ GeV which is the extrapolated value of the data of lattice simulations
for unquenched chiral quarks\cite{bowman05}. The same lattice data of Ref.\cite{bowman05}
are shown in the figure for comparison. The energy scale is fixed by taking $m=0.7$ GeV as suggested by the
study of ghost and gluon propagators. We could force the curve on the data by pushing the coupling towards
unreasonable large values since the agreement increases in Fig.\ref{C4} going from $\alpha_s=0.6$ ($M=0.94$) 
to $\alpha_s=1.8$ ($M=0.318$). That is probably a consequence of the decreasing of the mass parameter $M$.
Actually, in the curves of Fig.\ref{C4}  the mass function rises at an energy scale that is too large 
compared with the lattice data. In fact, a better agreement can be reached by a change of the overall scale,
as suggested by Fig.\ref{C5} where the mass function $M(p)$ is plotted for decreasing values of $m$ while keeping
$\alpha_s$ fixed at a reasonable value $\alpha_s=0.9$ ($M=0.57-0.65$). However, a scale $m$ smaller than
$0.2$ GeV would be in strong disagreement with the scale that emerges from the study of gluon and ghost
propagators. The inverse dressing function of the gluon, with its minimum, pinpoints the scale $m$ at the
value $m=0.73$ GeV for pure Yang-Mills SU(3)\cite{ptqcd2}. 
Including quarks, a slight increase at $m\approx 0.8$ was 
found in Section IV B while a slight decrease towards $m\approx 0.6$ was suggested by the ghost dressing
function in Section IV A. Thus an average $m\approx 0.7$ GeV seems to be a reasonable choice. On the other hand,
a 25\% reduction of the effective coupling was reported going from the quenched to the unquenched quark 
simulation\cite{twoloop}, so that a reasonable choice for the coupling seems to be $\alpha_s\approx 0.9$ in
the IR and we are left with the solid line in Fig.\ref{C4} and Fig.\ref{C5} ($m=0.7$ GeV, $M=0.65$ GeV, $\alpha_s=0.9$).

Quite interestingly, there is an other way to move the curve towards the data without forcing the parameters.
In Fig.\ref{C6} the mass function $M(p)$ is reported without changing the mass scales that are fixed at
$m=0.7$ GeV, $M=0.65$ GeV, but reducing the coupling $\alpha_s$ when the energy increases, thus simulating
the effect of a running coupling. The piecewise constant running coupling in Fig.\ref{C6} goes from a
large value $\alpha_s=0.9$ in the IR, down to $\alpha_s=0.3$ at $p_E\approx 2$ GeV, in good agreement
with the behavior of the effective coupling of lattice simulations\cite{twoloop,bogolubsky}.
While it is quite obvious that a change of the coupling is not justified in the 
present fixed-coupling expansion, the qualitative picture of Fig.~\ref{C6}
enforces the idea that a consistent RG improvement of the massive expansion would be mandatory
for a quantitative agreement with the lattice data at energies larger then the mass scales $m$, $M$.
A full RG running of masses and couplings was discussed in Ref.\cite{tissier14} for a massive phenomenological
model, with bare quark and gluon masses that were added to the Lagrangian.
The results are in qualitative agreement with the present work in the IR, but no quantitative comparison can
be made because of the different Lagrangians. In the framework of optimized perturbation theory a RG improved
expansion has been discussed in Ref.\cite{KN} for the chiral quark condensate. That work
could provide the natural framework for an extension of the present method.

The analytic functions $\sigma_p$, $\sigma_m$ in Eq.(\ref{sigmatot}) can be easily continued to
Minkowski space. The quark propagator can be written as
\BE
S(p)= S_p (p^2)\fsl{p}+S_M (p^2)
\label{SPM}
\EE
where the scalar functions $S_p$, $S_M$ follow from Eqs.(\ref{Sscalar}), (\ref{ZM2})
\begin{align}
S_M(p^2)&=\frac{Z(p)M(p)}{p^2-M(p)^2}=\frac{3M\sigma_0\sigma_M}{p^2(\sigma_0-\sigma_p)^2-9M^2\sigma_M^2}\nn\\
S_p(p^2)&=\frac{Z(p)}{p^2-M(p)^2}=\frac{\sigma_0(\sigma_0-\sigma_p)}{p^2(\sigma_0-\sigma_p)^2-9M^2\sigma_M^2}
\label{SPM2}
\end{align}
and here the functions $\sigma_p(x,s)$, $\sigma_M(x,s)$ are evaluated at $s=-p^2/m^2-i\varepsilon$.
The imaginary parts have a cut on the real positive axis $p^2>0$ where we can define two spectral densities
\begin{align}
\rho_M(p^2)&=-\frac{1}{\pi}\Im S_M(p^2)\nn\\
\rho_p(p^2)&=-\frac{1}{\pi}\Im S_p(p^2)
\label{rhoq}
\end{align}
so that the propagator reads
\BE
S(p)=\int_0^\infty {\rm d} q^2\frac{\rho_p(q^2) \fsl{p} + \rho_M (q^2)}{p^2-q^2+i\varepsilon}.
\label{spectral}
\EE
For a physical state, the K\"allen-Lehmann spectral densities contain important physical information
on the masses and on the thresholds of the multiparticle spectrum. Any observable fermion must satisfy
the positivity conditions\cite{IZ}
\BE
\rho_p(p^2)\geq 0
\label{cond1}
\EE
\BE
p\> \rho_p(p^2)-\rho_M(p^2)\geq 0
\label{cond2}
\EE
and the normalization condition
\BE
1=Z_0+\int_{q^2_0}^\infty {\rm d} q^2 \rho_p(q^2)
\label{norm}
\EE
where $q^2_0$ is the continuum threshold and $Z_0$ is the weight of the discrete one-particle state,
with $0\leq Z_0\leq 1$. Moreover the function $p\rho_p-\rho_M$ must have a support only on the continuum states and
the discrete one-particle term must cancel in the difference.

\begin{figure}[t] 
\centering
\includegraphics[width=0.35\textwidth,angle=-90]{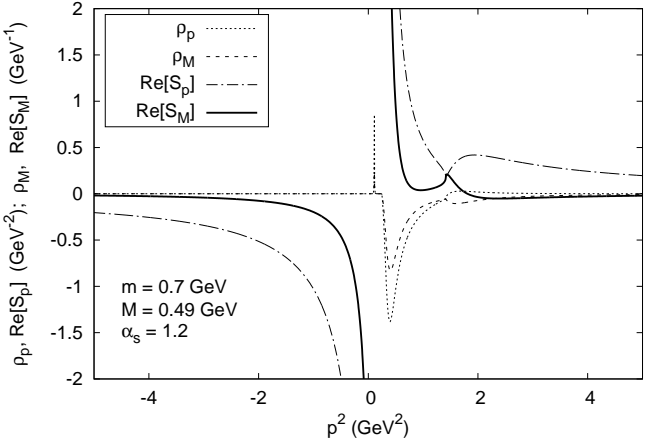}
\caption{The real parts $\Re S_M(p^2)$, $\Re S_p(p^2)$ of the quark propagator and the spectral
densities $\rho_M (p^2)$, $\rho_p (p^2)$ (imaginary parts) are shown as functions of the
physical momentum $p^2=-p_E^2$ for $m=0.7$ GeV, $\alpha_s=1.2$ and $M=0.49$ GeV.} 
\label{C7}
\end{figure}

\begin{figure}[t] 
\centering
\includegraphics[width=0.35\textwidth,angle=-90]{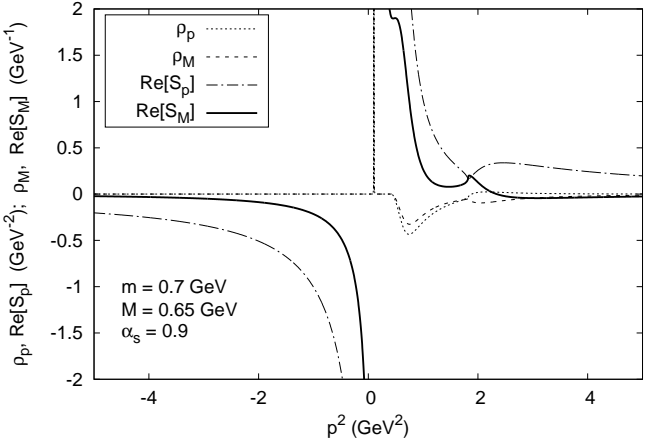}
\caption{The same as Fig.\ref{C7} but for $\alpha_s=0.9$ and $M=0.65$ GeV.}
\label{C8}
\end{figure}

\begin{figure}[t] 
\centering
\includegraphics[width=0.35\textwidth,angle=-90]{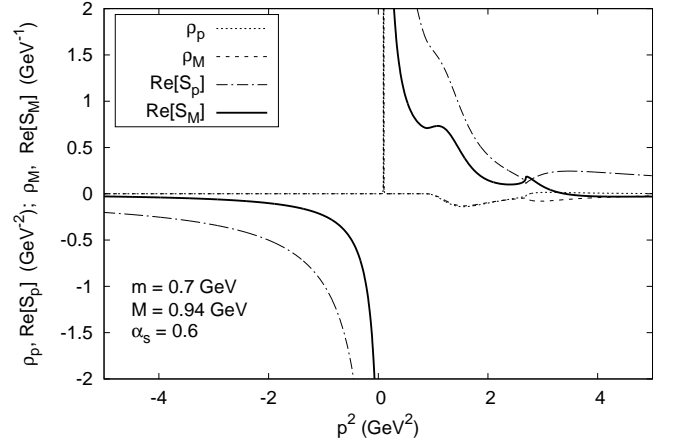}
\caption{The same as Fig.\ref{C7} but for $\alpha_s=0.6$ and $M=0.94$ GeV.}
\label{C9}
\end{figure}

\begin{figure}[!ht] 
\centering
\includegraphics[width=0.35\textwidth,angle=-90]{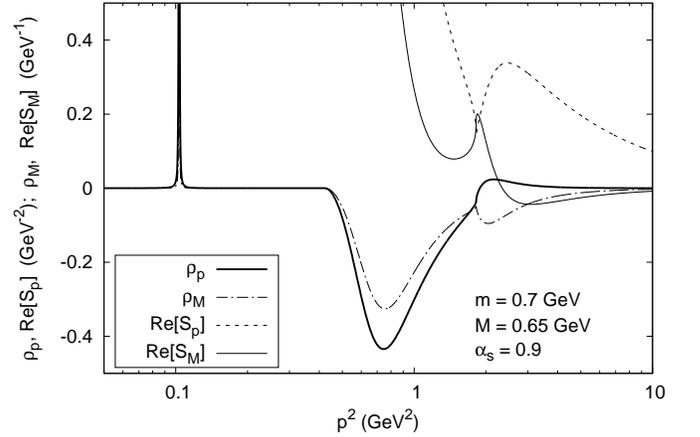}
\caption{Details of the quark spectral functions for $\alpha_s=0.9$, $M=0.65$ GeV, $m=0.7$ GeV
(same as Fig.\ref{C8}). The width of the peak at $p^2\approx 0.1$ depends on the finite value of
the imaginary part $\varepsilon$ in the analytic continuation $p_E^2=-p^2-i\varepsilon$. In the
limit $\varepsilon\to 0$ the peak becomes a genuine $\delta$-function and disappears from the plot.}
\label{C10}
\end{figure}

\begin{figure}[t] 
\centering
\includegraphics[width=0.35\textwidth,angle=-90]{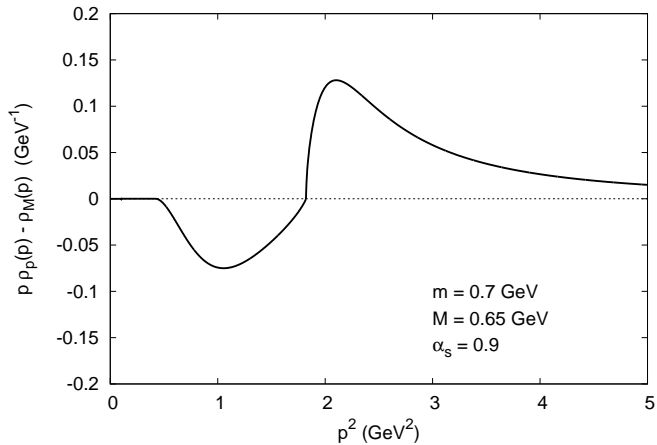}
\caption{The quark spectral function $[p\>\rho_p (p^2)-\rho_M(p^2)]$ is shown as a function of
the physical momentum $p^2=-p_E^2$ for $m=0.7$ GeV, $\alpha_s=0.9$ and $M=0.65$ GeV (same as Fig.\ref{C8}). 
The positivity condition of Eq.(\ref{cond2}) is violated for $p^2>q_1^2\approx M^2$ below the two-particle
threshold $q^2_2\approx (m+M)^2$.}
\label{C11}
\end{figure}

The real parts of the functions $S_p$, $S_M$ and the spectral functions $\rho_p$, $\rho_M$ are displayed in
Figs.\ref{C7},\ref{C8},\ref{C9} for $m=0.7$ GeV and $\alpha_s=1.2$, $0.9$, $0.6$, respectively.
The spectral functions are shown in more detail in Fig.\ref{C10} for $\alpha_s=0.9$, $M=0.65$ GeV. We recognize
a discrete term in the spectral functions at $p=M(0)\approx 0.32$ GeV, which arises from the pole of the
propagator. The finite width of the peak is just a measure of the finite value of $\varepsilon$ in the
numerical plot. Thus, a genuine discrete one-particle term $\rho_p,\rho_M\sim Z_0 \delta (p^2-M(0)^2)$ is
predicted by the exact calculation ($M(0)^2\approx (0.32)^2\approx 0.1$ GeV$^2$ in Fig.\ref{C10}). 
By a comparison of the plots we can identify two different thresholds.
A first threshold $q_1^2\approx M^2$ at the onset of a {\it negative} continuum spectral density 
($q_1^2\approx (0.65)^2\approx 0.42$ GeV$^2$ in Fig.\ref{C10}).
A second threshold $q_2^2\approx (M+m)^2$ where the spectral density turns positive
($q_2^2\approx (1.35)^2\approx 1.82$ GeV$^2$ in Fig.\ref{C10}). While this second threshold can be identified
with the usual two-particle threshold and the high-energy states have a
positive spectral density above $q_2\approx (m+M)$, the negative spectral density above $q_1\approx M$ has no obvious physical 
meaning. It violates the positivity condition (\ref{cond1}) and cannot be related to any kind of free-particle
behavior. Thus the quark propagator can only describe confined particles.

Moreover, if we look at the plot of the spectral function $[p\>\rho_p (p^2)-\rho_M(p^2)]$ in
Fig.\ref{C11}, we observe that the discrete term cancels in the difference and the spectral function
has a support only on the continuum states, above $q_1\approx M$, as predicted by general arguments\cite{IZ}.
But the positivity condition of Eq.(\ref{cond2}) is strongly violated for $q_1^2<p^2<q_2^2$. 
No complex poles are observed for the quark propagator and the analytic properties seem to be in agreement
with the qualitative predictions of Ref.\cite{alkofer}.

\section{Discussion}

Let us summarize the main findings of the paper. The massive expansion that was developed for pure Yang-Mills
theory in Refs.\cite{ptqcd,ptqcd2} has been extended to full QCD. While based on a perturbative expansion, the method
has a variational nature, with mass ratios and subtraction points that can be otpimized in order to minimize
the effects of higher loops. The method allows a unified description of the dynamical breaking of chiral
symmetry together with the dynamical gluon mass generation, from first principles and without adding any spurious
terms to the original Lagrangian. All mass divergences cancel exactly without the need for including other mass 
counterterms that would spoil the gauge and chiral symmetry of the theory. 
A finite result is achieved by the same wave function renormalization constants of the standard perturbation theory, 
thus ensuring that the correct behavior is predicted in the UV.

It is important to observe that the  mass does not arise as a trivial and direct consequence of using the massive 
propagators as an expansion point. In fact, that mass cancels at tree level because of the counterterms that leave
the Lagrangian unmodified. The mass emerges from the loops and is a genuine effect of the interactions.
For instance, it has been shown that no mass would emerge by the same method for the photon in QED.

The expansion is optimized by a comparison with the data of lattice simulations in the Euclidean space where an
impressive agreement is reached for the ghost and gluon sectors at one-loop\cite{scaling}. 
Some general universal scaling properties that where predicted for the dressing functions of pure Yang-Mills 
theory\cite{ptqcd,ptqcd2,scaling} are extended 
to full QCD where they are shown to be satisfied by the lattice data.

The limits of a fixed-coupling expansion emerge above 2 GeV and even below in the quark sector.
The problem could be cured by RG methods that become mandatory for a direct link between the IR and
the known phenomenology in the UV\cite{tissier14,KN}. 
Nevertheless, at a fixed coupling, a quark mass function emerges
that is in good qualitative agreement with the chiral limit of the lattice simulations.

At one-loop, explicit analytical functions are derived for the propagators. 
By a direct comparison with the lattice data,
the expansion is optimized in the Euclidean space yielding accurate analytic propagators that can
be easily continued to Minkowski space. Thus the method provides a powerful tool for the study of 
dynamical properties and spectral functions that can  be hardly extracted from any numerical
data set. From this point of view, the massive expansion is very predictive and gives
a direct proof of positivity violation and confinement for all the particles involved. 

In the gluon sector, the spectral function has no discrete
terms or real poles, in general agreement with previous numerical approaches\cite{dudal14}.
Moreover,  complex conjugated poles are found, in quantitative agreement with a fit\cite{sorella10} 
by the refined version\cite{dudal08,dudal08b,dudal11} of the Gribov-Zwanziger model\cite{GZ}.
The {\it i-particle} scenario\cite{iparticle} is confirmed and extended to full QCD where
two pairs of conjugated complex poles are found in the gluon propagator. That prediction is quite relevant
since no complex poles were found by the numerical calculation of Ref.\cite{straussDSE}.

A discrete one-particle term is found in the spectral function of the quark propagator at 
$p\approx M(0)\approx 0.32$ GeV. The propagator has no complex poles but positivity is badly
violated below the two-particle threshold.
While no direct dynamical content can be given to the gluon mass and to the mass parameters, the
discrete one-particle term can be identified as the (confined) physical mass of the constituent quarks.
On the other hand, the mass parameters $m$, $M$ are strongly related to the thresholds of the
spectral functions and determine their rich behavior that is observed in Minkowski space. Thus,
more than free-particle masses they have the physical meaning of threshold parameters.

Overall, while providing a powerful analytical tool and shading some light on the analytic properties 
of the propagators, the method can be improved in many ways. Among the many aspects that have not 
been addressed so far we mention a consistent use of RG equations\cite{tissier14,KN}, a variational estimate of 
the mass ratios by the effective potential (or by minimal sensitivity of physical 
observables\cite{stevenson,stevensonRS,stevenson16}) and  a formal extension to higher loops. The study
of other covariant gauges\cite{machado} would not be straightforward
or entirely analytical, but it would be very useful for addressing problems like the gauge dependence of the
propagators or the role played by the Abelian dominance in the maximally Abelian gauge\cite{abelian}.
Finally, while predicting complex poles for the gluon propagator, the present approach ignores the problem
of Gribov copies. Thus, it would be of some interest to explore effective models, like the Gribov-Zwanziger 
action, by the same optimized expansion.

\begin{widetext}

\appendix

\section{Explicit functions $F$ and $G$}

The functions $F(x)$ and $G(x)$ in Eq.(\ref{dress2}) were derived in Refs.\cite{ptqcd,ptqcd2}
by  the sum of polarization and self energy graphs for pure Yang-Mills theory up to one-loop and
third order.  The result is
\begin{align}
F(x)&=\frac{1}{x}+\frac{1}{72}\left[L_a+L_b+L_c+R_a+R_b+R_c\right]\nn\\
G(x)&=\frac{1}{12}\left[L_g+R_g\right]
\label{FGx}
\end{align}
where the logarithmic functions $L_x$ are
\begin{align}
L_a(x)&=\left[\frac{3x^3-34x^2-28x-24}{x}\right]\>
\sqrt{\frac{4+x}{x}}\>
\log\left(\frac{\sqrt{4+x}-\sqrt{x}}{\sqrt{4+x}+\sqrt{x}}\right)\nn\\
L_b(x)&=\frac{2(1+x)^2}{x^3}(3x^3-20x^2+11x-2)\log(1+x)\nn\\
L_c(x)&=(2-3x^2)\log(x)\nn\\
L_g(x)&=\frac{(1+x)^2(2x-1)}{x^2}\log(1+x)-2x\log(x)
\label{logsA}
\end{align}
and the rational parts $R_x$ are
\begin{align}
R_a(x)&=-\frac{4+x}{x}(x^2-20x+12)\nn\\
R_b(x)&=\frac{2(1+x)^2}{x^2}(x^2-10x+1)\nn\\
R_c(x)&=\frac{2}{x^2}+2-x^2\nn\\
R_g(x)&=\frac{1}{x}+2.
\label{rational}
\end{align}

\section{Standard one-loop self-energy}
The standard (uncrossed) one-loop quark self-energy arises from the first self-energy graph in Fig.\ref{F1}.
It is the usual fermionic self-energy in Landau gauge, but a mass is included in the
gluon propagator $\Delta_m$ inside the loop, according to Eq.(\ref{Deltam}):
\BE
\Sigma^{(1)}(p)=C_f g^2\int \kkki\left\{
\frac{\gamma^\mu t_{\mu\nu}(k)(\fsl{p}+\fsl{k}+M)\gamma^\nu}
{\left[-k^2+m^2\right]\left[(p+k)^2-M^2\right]}
\right\}.
\EE
By some algebra of gamma matrices and switching to Euclidean space, 
using $\fsl{p}=i \fsl{p_E}$ and $\{\gamma_E^\mu,\gamma_E^\nu\}=2\delta^{\mu\nu}$,
\BE
\Sigma^{(1)}(p)=C_f g^2\int \kkkE\left\{
\frac{i \fsl{p_E}-3M+i \fsl{k_E}\left[3+\displaystyle{ \frac{2(p_E\cdot k_E)}{k_E^2} }\right]} 
{\left[k_E^2+m^2\right]\left[-(p_E+k_E)^2-M^2\right]}
\right\}
\EE
and the self-energy can be written as
\BE
\Sigma^{(1)}(p)=\frac{C_f g^2}{16\pi^2}\left\{
(3M-i\fsl{p_E})I_1(p)-3i\gamma_E^\mu I_2^\mu(p)+\frac{2i\gamma_E^\mu p_E^\nu}{m^2} 
\left[I_3^{\mu\nu}(p,m)-I_3^{\mu\nu}(p,0)\right] \right\}
\label{S1}
\EE
where the integrals $I_1$, $I_2^\mu$, $I_3^{\mu\nu}$ can be written in terms of Passarino-Veltman
functions 
\BE
I_1(p)=B_0(p,m,M)=(16\pi^2)\int\kkEd \frac{1}{\left[k_E^2+m^2\right]\left[(p_E+k_E)^2+M^2\right]}
\EE
\BE
I_2^\mu(p)=p_E^{\>\mu} B_1(p,m,M)=(16\pi^2)\int\kkEd \frac{k_E^\mu}{\left[k_E^2+m^2\right]\left[(p_E+k_E)^2+M^2\right]}
\EE
\BE
I_3^{\mu\nu} (p,m)=p_E^{\>\mu}p_E^{\>\nu}\> B_{21}(p,m,M)+\delta^{\mu\nu} B_{22}(p,m,M)
=(16\pi^2)\int\kkEd \frac{k_E^\mu k_E^\nu}{\left[k_E^2+m^2\right]\left[(p_E+k_E)^2+M^2\right]}
\EE
and can be evaluated by dimensional regularization, setting $d=4-\epsilon$.
In units of $m$, denoting by $s=p_E^2/m^2$ the Euclidean momentum and by $x=M^2/m^2$ the squared quark mass,
the finite parts read
\BE
B_0^f(x,s)=2+\frac{t}{s}\log R-\frac{t+s-x+1}{2s}\>\log x
\EE
\BE
B_1^f(x,s)=-1+\frac{1-y_-^2}{2}\log x -\frac{x-1}{2s}-\frac{t\>w}{2s^2} \> \log R
\EE
\BE
B_{21}^f(x,s)=\frac{7}{18}+\frac{y_-^3-1}{3}\log x +\frac{x-1}{6s}+\frac{w^2+2s}{3s^2}+
\frac{t(w^2+s)}{3s^3}\log R
\EE
\BE
B_{22}^f(x,s)=-\frac{m^2}{2}\left[B_0^f(x,s)-(s+x-1)B_1^f(x,s)
-s B_{21}^f(x,s)\right]
\label{B22}
\EE
where
\BE
w=s+x-1,\qquad t=\sqrt{w^2+4s}=\sqrt{(s+x)^2+2(s-x)+1}
\label{kt}
\EE
and
\BE
y_{\pm}=\frac{w\pm t}{2s}=\frac{\pm t+s+x-1}{2s}
\label{ypm}
\EE
are the positive and negative solutions of the equation
\BE
-sy^2+wy+1=s(y-y_-)(y_+ -y)=0.
\EE
By setting $y=1$, they are easily shown to satisfy the identities
\BE
(1-y_-)(y_+ -1)=\frac{x}{s},\quad y_- y_+=-\frac{1}{s}
\EE
so that the positive ratio $R$ can be defined as
\BE
R=\frac{(y_+ -1)}{y_+}=\frac{-xy_-}{1-y_-}
\EE
and we can write it explicitly as
\BE
R=\frac{t-s+x-1}{t+s+x-1}=x\>\frac{t-s-x+1}{t+s-x+1}
\label{R}
\EE
yielding
\BE
\log\left[\frac{R}{\sqrt{x}}\right]=\frac{1}{2}\log\left\{\frac{(t-s)^2-(x-1)^2}{(t+s)^2-(x-1)^2}\right\}.
\EE
The diverging parts are
\BE
B_0^\epsilon (x,s)=\frac{2}{\epsilon}+\log \frac{\mu^2}{m^2}
\EE
\BE
B_1^\epsilon (x,s)=-\frac{1}{2}\left[\frac{2}{\epsilon}+\log \frac{\mu^2}{m^2}\right]
\EE
\BE
B_{21}^\epsilon (x,s)=\frac{1}{3}\left[\frac{2}{\epsilon}+\log \frac{\mu^2}{m^2}\right]
\label{B21e}
\EE
\BE
B_{22}^\epsilon (x,s)=-\frac{m^2}{4}\left(\frac{s}{3}+x+1\right)
\left[\frac{2}{\epsilon}+\log \frac{\mu^2}{m^2}+1\right].
\label{B22e}
\EE
In Eq.(\ref{S1}) we also need the subtracted functions $\Delta B_X(p,m,M)=B_X (p,m,M)-B_X (p,0,M)$.
In the limit $m\to 0$ the products $m^2 t$, $m^2 w$ are finite and Eqs.(\ref{kt}),(\ref{ypm}),(\ref{R}) give
\begin{align}
\lim_{m\to 0} (m^2 t)&=\lim_{m\to 0} (m^2 w)=M^2+p_E^2\nn\\
\lim_{m\to 0} R&=\frac{M^2}{M^2+p_E^2}=\frac{x}{s+x}\nn\\
\lim_{m\to 0} y_-&=0
\end{align}
so that, inserting an IR cutoff $m_0\to 0$, the finite parts of the subtracted functions read
\BE
\Delta B_0^f(x,s)=\left\{\frac{t}{s}\log R-\frac{t-s-x+1}{2s}\>\log x-\frac{s+x}{s}\log\frac{x}{x+s}\right\}
+\log\frac{m^2}{m_0^2}
\label{DB0}
\EE
\BE
\Delta B_1^f(x,s)= \left\{\frac{1}{2s}-\frac{t\>w}{2s^2} \> \log R-\frac{y_-^2}{2}\log x
+\frac{(s+x)^2}{2s^2}\log\frac{x}{x+s}\right\}
-\frac{1}{2}\log\frac{m^2}{m_0^2}
\label{DB1}
\EE
\BE
\Delta B_{21}^f(x,s)=\left\{\frac{y_-^3}{3}\log x-\frac{1}{6s}+\frac{1-2x}{3s^2}+
\frac{t(w^2+s)}{3s^3}\log R-\frac{(s+x)^3}{3s^3}\log\frac{x}{x+s}\right\}
+\frac{1}{3}\log\frac{m^2}{m_0^2}.
\label{DB21}
\EE
When subtracted, the diverging part of the dimensionless function $B_{21}$ in Eq.(\ref{B21e})
is finite in the UV and cancels the IR divergence of the finite part in Eq.(\ref{DB21})
\BE
\Delta B_{21}^\epsilon (x,s)=-\frac{1}{3}\log\frac{m^2}{m_0^2},
\EE
while the diverging part $B^\epsilon_{22}$ in Eq.(\ref{B22e}), which is of mass dimension 2,
gives the diverging term
\BE
\Delta B_{22}^\epsilon (x,s)=-\frac{m^2}{4}\left\{
\left(\frac{2}{\epsilon}+\log \frac{\mu^2}{m^2}+1\right)
-\left(\frac{s}{3}+x\right)\log \frac{m^2}{m_0^2}\right\}.
\label{DB22e}
\EE 
The finite part in Eq.(\ref{B22}), when subtracted, gives
\BE
\Delta B_{22}^f (x,s)=-\frac{m^2}{2}\left[B_0^f(x,s)+B_1^f(x,s)-(s+x)\>\Delta B_1^f(x,s)
-s\>\Delta B_{21}^f(x,s)\right].
\EE
Inserting Eqs.(\ref{DB1}-\ref{DB21}), we observe that the IR diverging terms in $\Delta B^f_{22}$
cancel the IR diverging term of $\Delta B_{22}^\epsilon$ in Eq.(\ref{DB22e}), so that no term depending
on $m_0$ survives and we can safely drop them everywhere.

Inserting these results into Eq.(\ref{S1}), taking $C_f=4/3$ and restoring $\fsl{p}=i \fsl{p_E}$, 
the one-loop self energy can be written as
\BE
\Sigma^{(1)}(p)=\Sigma^{(1)}_M (p)+\fsl{p} \>\Sigma^{(1)}_p (p)
\EE
where
\BE
\Sigma^{(1)}_M (p)=\left(\frac{\alpha_s}{\pi}\right) M \>\sigma_M (x,s), \qquad
\Sigma^{(1)}_p (p)=\left(\frac{\alpha_s}{3 \pi}\right) \>\sigma_p \>(x,s)
\EE
and the adimensional functions $\sigma_M$, $\sigma_p$ read
\begin{align}
\sigma_M (x,s)&=B_0^f (x,s)+B_0^\epsilon (x,s)\nn\\
\sigma_p (x,s)&=-B_0^f (x,s)-3B_1^f (x,s)+\frac{2\>\Delta B_{22}^f (x,s)}{m^2}+2s\>\Delta B_{21}^f (x,s)
+\sigma_p^\epsilon
\label{sig1}
\end{align}
where the constant term $\sigma_p^\epsilon$ arises from the UV diverging parts 
\BE
\sigma_p^\epsilon=-B_0^\epsilon-3 B_1^\epsilon+\frac{2}{m^2} \>\Delta B_{22}^\epsilon=-\frac{1}{2}
\EE
and is finite
so that the divergences cancel entirely in the function $\Sigma_p^{(1)}$ which is finite. 
Summing up all terms we find
\begin{align}
\sigma_M (x,s)&=\left\{\frac{t}{s}\log R-\frac{t+s-x+1}{2s}\>\log x\right\}
+\left\{\frac{2}{\epsilon}+\log \frac{\mu^2}{m^2}\right\}\nn\\
\sigma_p (x,s)&=C_R\>\log R+C_{x}\> \log x+ C_{xs}\> \log\frac{x}{x+s} + C_0
\label{sigma1}
\end{align}
where the functions $C_R$, $C_{x}$, $C_{xs}$, $C_0$ read
\begin{align}
C_R&=\frac {t}{2s^2}\left[(x+s)^2+(x-s)-2\right]\nn\\
C_{x}&=-\frac{1}{2} C_R+\frac {1}{4s^2}\left[(x+s)^3-3(x-s)+2\right]\nn\\
C_{xs}&=-\frac {(x+s)^3}{2s^2}\nn\\
C_0&=\frac{x-2}{2s}-\frac{1}{2}.
\label{CC}
\end{align}
The explicit result coincides with that reported by other authors\cite{tissier14} up to a scheme-dependent
irrelevant additive constant in the last term, which is usually absorbed by a finite wave-function renormalization.
It can be checked
that $\sigma_p$ becomes a constant in the limit $m^2\to 0$, 
which is a well known result of the standard perturbative expansion.

\section{Crossed self-energy graphs}

The crossed one-loop graphs in Fig.\ref{F1} are obtained by insertion of a counterterm in the gluon and
quark lines. The explicit expressions can be easily recovered by a derivative according to the
general Eq.(\ref{crossedsum}).

The insertion of the gluon mass counterterm $\delta\Gamma_g=m^2$ in the standard one-loop self energy
gives the first crossed self-energy graph in Fig.\ref{F1}
\BE
\Sigma^{g} (p)=-m^2\frac{\partial }{\partial m^2}\Sigma^{(1)} (p)
\label{crg}
\EE
The second crossed self-energy graph in Fig.\ref{F1} is obtained by insertion of the quark mass counterterm
$\delta\Gamma_q=-M$ 
\BE
\Sigma^{q} (p)=-M\frac{\partial}{\partial M} \Sigma^{(1)} (p).
\label{crq}
\EE
With the notation of Appendix B, the sum of all three one-loop graphs in Fig.\ref{F1} gives
\BE
\Sigma (p)=\Sigma^{(1)} (p) +\Sigma^{g} (p)+\Sigma^{q} (p)=
\left(\frac{\alpha_s}{\pi}\right) M \> \sigma_M (x,s)
+\fsl{p} \>\left(\frac{\alpha_s}{3 \pi}\right) \>\sigma_p \>(x,s)
\EE
where
\begin{align}
\sigma_M (x,s)&=\sigma_M^1 (x,s)+\sigma_M^g (x,s)+\sigma_M^q (x,s)\nn\\
\sigma_p (x,s)&=\sigma_p^1 (x,s)+\sigma_p^g (x,s)+\sigma_p^q (x,s)
\label{sigsum}
\end{align}
and $\sigma_M^1$, $\sigma_p^1$ are the adimensional functions that were recovered
for the standard (uncrossed) one-loop graph in Eq.(\ref{sigma1}) of Appendix B.
The insertion of the gluon counterterm $\delta \Gamma_g$ gives the crossed functions 
$\sigma_M^g$, $\sigma_p^g$ that by Eq.(\ref{crg}) can be written
\begin{align}
\sigma_M ^g(x,s)&= 1+\left(s\frac{\partial}{\partial s}+x\frac{\partial}{\partial x}\right) \sigma_M^1 (x,s)\nn\\
\sigma_p ^g(x,s)&= \left(s\frac{\partial}{\partial s}+x\frac{\partial}{\partial x}\right) \sigma_p^1 (x,s).
\label{sigmag}
\end{align}
The insertion of the quark counterterm $\delta \Gamma_q$ gives the crossed functions 
$\sigma_M^q$, $\sigma_p^q$ that by Eq.(\ref{crq}) can be written
\begin{align}
\sigma_M ^q(x,s)&= \left(-1-2x\frac{\partial}{\partial x}\right) \sigma_M^1 (x,s)\nn\\
\sigma_p ^q(x,s)&= -2x\frac{\partial}{\partial x} \sigma_p^1 (x,s).
\label{sigmaq}
\end{align}

For future reference, let us first evaluate the sum of the first two terms that would be useful for
a theory without quark counterterms (for instance, when the chiral symmetry is broken by explicit
bare mass terms in the Lagrangian). Inserting Eq.(\ref{sigma1}) in Eq.(\ref{sigmag}) we find
\begin{align}
\sigma_M^1(x,s)+\sigma_M ^g(x,s)&=
\left\{\frac{(s+x)^2+(s-x)}{st}\log \frac{R}{\sqrt{x}}-\frac{s-x}{2s}\>\log x\right\}
+\left\{\frac{2}{\epsilon}+\log \frac{\mu^2}{m^2}\right\}\nn\\
\sigma_p^1(x,s)+\sigma_p ^g(x,s)&=C^g_R\>\log R+C^g_{x}\> \log x+ C^g_{xs}\> \log\frac{x}{x+s} + C^g_0
\label{sigma1g}
\end{align}
where the new functions $C^g_R$, $C^g_{x}$, $C^g_{xs}$, $C^g_0$ read
\begin{align}
C^g_R&=\frac {1}{s^2t}\left[(x+s)^4+(s-x)(s+x)^2+2sx+(s-x)+1\right]\nn\\
C^g_{x}&=-\frac{1}{2} C^g_R+\frac {1}{2s^2}\left[(x+s)^3-1\right]\nn\\
C^g_{xs}&=-\frac {(x+s)^3}{s^2}\nn\\
C^g_0&=\frac{1+x}{s}.
\label{CCg}
\end{align}
The sum $\sigma_M^1+\sigma_M^g$ is still divergent but the divergence can be canceled by the bare mass
of the quark, while only a constant would be added to the last term by a finite 
wave-function renormalization.

In the chiral limit there is no  bare mass for the quark. However, the divergence is canceled by the third crossed
graph. Adding up all three graphs, by Eqs.(\ref{sigsum}), (\ref{sigmag}), (\ref{sigmaq}), we can write
\begin{align}
\sigma_M (x,s)&= 1+\left(s\frac{\partial}{\partial s}-x\frac{\partial}{\partial x}\right) \sigma_M^1 (x,s)\nn\\
\sigma_p ^g(x,s)&= \left(1+s\frac{\partial}{\partial s}-x\frac{\partial}{\partial x}\right) \sigma_p^1 (x,s),
\label{sigtot}
\end{align}
and inserting Eq.(\ref{sigma1}) we find the following explicit expressions
\begin{align}
\sigma_M (x,s)&=
\left\{\frac{1-2x}{2s}\>\log x-\frac{s(2x+1)+(2x-1)(x-1)}{st}\log \frac{R}{\sqrt{x}}\right\}\nn\\
\sigma_p \>(x,s)&=C^{gq}_R\>\log R+C^{gq}_{x}\> \log x+ C^{gq}_{xs}\> \log\frac{x}{x+s} + C^{gq}_0
\label{sigmatot}
\end{align}
where the complete functions $C^{gq}_R$, $C^{gq}_{x}$, $C^{gq}_{xs}$, $C^{gq}_0$ read
\begin{align}
C^{gq}_R&=\frac {1}{s^2t}\left\{
(s-2x)\left[(x+s)^3+(s^2-x^2)\right]+(s-x+1)(1-3x)+2sx\right\}\nn\\
C^{gq}_{x}&=-\frac{1}{2} C^{gq}_R+\frac {1}{2s^2}\left[(x+s)^2 (s-2x)+3x-1\right]\nn\\
C^{gq}_{xs}&=\frac{(2x-s)(x+s)^2}{s^2}\nn\\
C^{gq}_0&=\frac{1-2x}{s}+1.
\label{CCtot}
\end{align}
Here the additive constant in the last term has been set by requiring that $\sigma_p\to 0$ in the limit
$s\to \infty$. That is equivalent to taking $\delta Z_q=0$ at the special subtraction point $\mu\to\infty$ 
in Eq.(\ref{mudef}).
The total finite functions $\sigma_M$, $\sigma_p$ in Eq.(\ref{sigmatot}) are used in Eqs.(\ref{sigmadef}),
(\ref{ZM2}) of Section IV B.
In the UV, for $s\to\infty$, the function $\sigma_M$ tends to zero as $\log s/s$ according to 
the asymptotic behavior
\BE
\sigma_M (x,s) \approx \frac{1}{s}\left[(1+2x)\log s-2x\log x\right].
\label{asymptM}
\EE
In the IR, $\sigma_M$ is a finite analytic function of $s$ and its first order expansion is
\BE
\sigma_M (x,s) \approx \left[\frac{(2x-1)(x-1)-x\log x}{(x-1)^2}\right]
-s\left[\frac{(2x^2+5x-1)(x-1)-6x^2\log x}{2(x-1)^4}\right]
\label{taylorM}.
\EE
It reaches a finite value at $s=0$ and decreases for $s>0$.

\end{widetext}


\begin{thebibliography} {99}

\bibitem{cornwall} J. M. Cornwall, Phys. Rev. D 26, 1453 (1982)

\bibitem{bogolubsky} I.L. Bogolubsky, E.M. Ilgenfritz, M. Muller-Preussker, A. Sternbeckc, 
Phys. Lett. B {\bf 676}, 69 (2009).
\bibitem{twoloop}  E.-M. Ilgenfritz, M. M\"uller-Preussker, A. Sternbeck, A. Schiller, arXiv:hep-lat/0601027.
\bibitem{dudal} D. Dudal, O. Oliveira, N. Vandersickel, Phys. Rev. D {\bf 81}, 074505 (2010).
\bibitem{binosi12} A. Ayala, A. Bashir, D. Binosi, M. Cristoforetti and J. Rodriguez-Quintero, 
Phys. Rev. D {\bf 86}, 074512 (2012)
\bibitem{burgio15} G. Burgio, M. Quandt, H. Reinhardt, H. Vogt, Phys. Rev. D 92, 034518 (2015).
\bibitem{bowman04} P. O. Bowman, U. M. Heller, D. B. Leinweber, M. B. Parappilly, A. G. Williams, 
Phys. Rev. D {\bf 70}, 034509 (2004).
\bibitem{bowman05} P. O. Bowman, U. M. Heller, D. B. Leinweber, M. B. Parappilly, A. G. Williams, J. Zhang, 
Phys. Rev. D {\bf 71}, 054507 (2005).
\bibitem{su2glu} V. G. Bornyakov, V. K. Mitrjushkin, M. M\"ller-Preussker, Phys. Rev. D {\bf 81}, 054503 (2010).
\bibitem{su2gho} V. G. Bornyakov, E.-M. Ilgenfritz, C. Litwinski, 
V. K. Mitrjushkin, M. M\"uller-Preussker, Phys.Rev.D {\bf 92}, 074505 (2015), arXiv:1302.5943.

\bibitem{aguilar8} A. C. Aguilar, D. Binosi, J. Papavassiliou, Phys. Rev. D78, 025010 (2008).
\bibitem{aguilar10} A. C. Aguilar, J. Papavassiliou, Phys. Rev. D81, 034003 (2010).
\bibitem{aguilar14} A. C. Aguilar, D. Binosi, J. Papavassiliou, Phys. Rev. D 89, 085032 (2014). 
\bibitem{aguilar14b} A. C. Aguilar, D. Binosi, D. Ibanez, J. Papavassiliou, Phys.Rev.D {\bf 90}, 065027 (2014), arXiv:1405.3506. 

\bibitem{papa15} A. C. Aguilar, D. Binosi, J. Papavassiliou, Phys. Rev. D {\bf 91}, 085014 (2015).
\bibitem{papa15b} D. Binosi, L. Chang, J. Papavassiliou, C. D. Roberts, Phys. Lett. B {\bf 742}, 183 (2015).

\bibitem{huber14} A. L. Blum, M. Q. Huber, M. Mitter, L. von Smekal, Phys. Rev. D {\bf 89}, 061703 (2014).
\bibitem{huber15g} M. Q. Huber, Phys. Rev. D 91, 085018 (2015).
\bibitem{huber15b} A. K. Cyrol, M. Q. Huber, L. von Smekal, Eur.Phys.J. C75 (2015) 102.

\bibitem{fischer2009} C. S. Fischer, A. Maas, J. M. Pawlowski, Annals Phys. {\bf 324}, 2408 (2009).
\bibitem{fischer2003} C. S. Fischer and R. Alkofer, Phys.Rev. D {\bf 67}, 094020 (2003). 

\bibitem{reinhardt04} C. Feuchter and H. Reinhardt, Phys. Rev. D {\bf 70}, 105021 (2004).
\bibitem{reinhardt05} H. Reinhardt and C. Feuchter, Phys.Rev. D 71,  105002, (2005).
\bibitem{reinhardt08} D. Epple, H. Reinhardt, W. Schleifenbaum, A.P. Szczepaniak, Phys. Rev. D77, 085007,(2008).
\bibitem{reinhardt14} M. Quandt, H. Reinhardt, J. Heffner, Phys. Rev. D 89, 065037 (2014).

\bibitem{sigma} F. Siringo and L. Marotta, Eur. Phys. J. C {\bf  44}, 293 (2005).
\bibitem{sigma2} F. Siringo, Mod. Phys. Lett. A {\bf 29}, 1450026 (2014), arXiv:1308.4037 
\bibitem{gep2}  F. Siringo, Phys. Rev. D {\bf 88}, 056020 (2013), arXiv:1308.1836.
\bibitem{varqed} F. Siringo, Phys. Rev. D {\bf 89},  025005 (2014), arXiv:1308.2913.
\bibitem{varqcd} F. Siringo, Phys. Rev. D {\bf  90}, 094021 (2014), arXiv:1408.5313.
\bibitem{genself} F. Siringo, Phys. Rev. D {\bf 92}, 074034 (2015), arXiv:1507.00122.
\bibitem{ptqcd0} F. Siringo, {\it Perturbation theory of non-perturbative QCD}, arXiv:1507.05543.

\bibitem{straussDSE} S. Strauss, C. S. Fischer, C. Kellermann, Phys. Rev. Lett. {\bf 109}, 252001 (2012).
\bibitem{dudal14} D. Dudal, O. Oliveira, P. J. Silva, Phys. Rev. D {\bf 89}, 014010 (2014).

\bibitem{shakinPRD} X. Li, C. M. Shakin, Phys. Rev. D {\bf 71} 074007 (2005).

\bibitem{iparticle} L. Baulieu, D. Dudal, M. S. Guimaraes, M. Q. Huber,
S. P. Sorella, N. Vandersickel, D. Zwanziger, Phys.Rev.D {\bf 82}, 025021 (2010).

\bibitem{tissier10} M. Tissier, N. Wschebor, Phys. Rev. D {\bf 82}, 101701(R) (2010).
\bibitem{tissier11} M. Tissier, N. Wschebor, Phys. Rev. D {\bf 84}, 045018 (2011).
\bibitem{tissier14} M. Pelaez, M. Tissier, N. Wschebor, Phys. Rev. D {\bf 90}, 065031 (2014).

\bibitem{sorella15} M.A.L. Capri, A.D. Pereira, R.F. Sobreiro, S.P. Sorella, Eur. Phys. J. C {\bf 75}, 479 (2015).  
\bibitem{dudal15} M.A.L. Capri, D. Dudal, D. Fiorentini, M.S. Guimaraes, I.F. Justo, A.D. Pereira, 
B.W. Mintz, L.F. Palhares, R.F. Sobreiro, S.P. Sorella, Phys. Rev. D {\bf 92}, 045039 (2015). 
\bibitem{dudal08} D. Dudal, J. A. Gracey, S. P. Sorella, N. Vandersickel, H. Verschelde, Phys. Rev. D {\bf 78},
065047 (2008).
\bibitem{dudal08b} D.Dudal, S.P.Sorella, N.Vandersickel, H.Verschelde, Phys. Rev. D {\bf 77}, 071501 (2008).
\bibitem{dudal11} D. Dudal, S. P. Sorella, N. Vandersickel, Phys. Rev. D {\bf 84}, 065039 (2011).

\bibitem{GZ} D. Zwanziger, Nucl. Phys. B {\bf 323}, 513 (1989).

\bibitem{ptqcd}  F. Siringo, {\it Perturbative study of Yang-Mills theory in the infrared}, arXiv:1509.05891.
\bibitem{ptqcd2} F. Siringo, Nucl. Phys. B {\bf 907}, 572 (2016), [arXiv:1511.01015].
\bibitem{scaling} F. Siringo, {\it Universal scaling of gluon and ghost propagators in the infrared}, arXiv:1607.02040.

\bibitem{stevenson} P.M. Stevenson, Phys. Rev. D {\bf 32}, 1389 (1985).
\bibitem{stevensonRS} P.M. Stevenson, Nucl. Phys. B {\bf 868}, 38 (2013).
\bibitem{stevenson16} P.M. Stevenson, {\it Exploring arbitrarily high orders of optimized perturbation theory 
in QCD with nf -> 16.5}, Nucl.Phys.B {\bf 910}, 469 (2016), arXiv:1606.09500.
\bibitem{KN1} J.-L. Kneur and A. Neveu, Phys. Rev. D {\bf 81}, 125012 (2010); 
\bibitem{KN2} J.-L. Kneur and A. Neveu, Phys. Rev. D {\bf 85}, 014005 (2012).
\bibitem{KN3} J.-L. Kneur and A. Neveu, Phys. Rev. D {\bf 88}, 074025 (2013). 
\bibitem{KN}  J.-L. Kneur and A. Neveu, Phys. Rev. D {\bf 92}, 074027 (2015).

\bibitem{var} F. Siringo,  Phys. Rev. D {\bf 62}, 116009 (2000).
\bibitem{light} F. Siringo, Europhys. Lett. {\bf 59}, 820 (2002).
\bibitem{bubble} F. Siringo and L. Marotta, Int. J. Mod. Phys. {\bf A25}, 5865 (2010), arXiv:0901.2418v2.

\bibitem{su2} F. Siringo, L. Marotta, Phys. Rev. D {\bf 78}, 016003 (2008).
\bibitem{LR} F. Siringo and L. Marotta, Phys. Rev. D {\bf 74}, 115001 (2006).
\bibitem{HT} F. Siringo, Phys. Rev. D {\bf 86}, 076016 (2012), arXiv: 1208.3592v2.
\bibitem{stancu2} I. Stancu and P. M. Stevenson, Phys. Rev. D {\bf 42}, 2710 (1990).
\bibitem{stancu} I. Stancu, Phys. Rev. D {\bf 43}, 1283 (1991).
\bibitem{superc1} M. Camarda, G.G.N. Angilella, R. Pucci, F. Siringo, 
Eur. Phys. J. B {\bf 33}, 273 (2003).
\bibitem{superc2} L. Marotta, M. Camarda, G.G.N. Angilella and F. Siringo, 
Phys. Rev. B {\bf 73}, 104517 (2006).             
\bibitem{AF} L. Marotta and F. Siringo, Mod. Phys. Lett. B, {\bf 26}, 1250130 (2012), arXiv:0806.4569v3.

\bibitem{OZ} R, Oehme and W. Zimmermann, Phys. Rev. D {\bf 21}, 471 (1980).

\bibitem{sorella10} S. P. Sorella, D. Dudal, M. S. Guimaraes, and N. Vandersickel, 
Proceedings, Workshop on The many faces of QCD (FacesQCD2010), PoS FACESQCD, 022 (2010), arXiv:1102.0574.

\bibitem{kondo} K.-I. Kondo, Phys.Lett. B {\bf 514}, 335 (2001).

\bibitem{taylor} J. C. Taylor, Nucl. Phys. B {\bf 33}, 436 (1971). 

\bibitem{IZ} C. Itzykson, J.-B. Zuber, {\it Quantum Field Theory}, McGraw-Hill (1980).

\bibitem{alkofer} R. Alkofer, W. Detmold, C. S. Fischer and P. Maris, Phys.Rev.D {\bf 70}, 014014 (2004).

\bibitem{machado} F. A. Machado, {\it Transversality of gluon mass generation through an effective 
loop expansion in covariant and background field gauges}, arXiv:1601.02067.

\bibitem{abelian} A. S. Kronfeld, G. Schierholz, and U.-J. Wiese, Nucl. Phys. B {\bf 293}, 461 (1987); A. S. Kronfeld,
M. L. Laursen, G. Schierholz, and U.-J. Wiese, Phys. Lett. B {\bf 198}, 516 (1987).

\end{thebibliography}
\end{document}